\documentclass[11pt, a4paper]{article}
\usepackage[top=0.80in, bottom=0.80in, left=1.00in, right=1.00in]{geometry}
\usepackage{booktabs}
\usepackage{multirow}
\usepackage{setspace}
\usepackage{natbib}
\usepackage{authblk}
\usepackage[pdftex]{graphicx}
\usepackage{amsmath}
\usepackage{amsthm}
\usepackage{mathtools}
\usepackage{amsfonts}
\usepackage{ascmac}
\usepackage{amssymb}
\usepackage{bm}
\usepackage{color}
\usepackage{enumitem}
\usepackage{mathrsfs}
\usepackage{lscape}
\usepackage[
	colorlinks = true, 
	anchorcolor=black,
	linkcolor = blue,
	citecolor = blue, 
	filecolor  = black, 
	urlcolor   = blue
]{hyperref}
\usepackage{comment}
\usepackage{algorithmic, algorithm}
\usepackage{refcount}

\usepackage{longtable}
\usepackage[font=normalsize]{caption}     
\usepackage{etoolbox}
\AtBeginEnvironment{longtabu}{\footnotesize}{}{}   

\usepackage{breqn}
\allowdisplaybreaks[1]

\newcommand{\biblist}{\begin{list}{}
{\listparindent 0.0cm \leftmargin 0.50cm \itemindent -0.50 cm
\labelwidth 0 cm \labelsep 0.50 cm
\usecounter{list}}\clubpenalty4000\widowpenalty4000}
\newcommand{\ebiblist}{\end{list}}

 \theoremstyle{plain}
\newtheorem{thm}{Theorem}

\theoremstyle{definition}

\theoremstyle{remark}
\newtheorem*{rem}{Remark}

\newcommand{\ep}{\varepsilon}

\newcommand{\pd}{\partial}

\newcommand{\E}{\mathbb{E}}

\newcommand{\argmin}{\mathop{\rm arg~min}\limits}

\title{\bf Robustness against outliers \\in ordinal response model via divergence approach}
\author{Tomotaka Momozaki}
\author{Tomoyuki Nakagawa}
\affil{Tokyo University of Science}

\date{Last update: \today}

\begin{document}
\maketitle
\begin{abstract}
This study deals with the problem of outliers in ordinal response model, which is a regression on ordered categorical data as the response variable. 
``Outlier" means that the combination of ordered categorical data and its covariates is heterogeneous compared to other pairs. 
Although the ordinal response model is important for data analysis in various fields such as medicine and social sciences, it is known that the maximum likelihood method with probit, logit, log-log and complementary log-log link functions, which are often used, is strongly affected by outliers, and statistical analysts are forced to limit their analysis when there may be outliers in the data.

To solve this problem, this paper provides inference methods with two robust divergences (the density-power and $\gamma$-divergences).
We also derive influence functions for the proposed methods and show conditions on the link function for them to be bounded and to redescendence.
Since the commonly used link functions satisfy these conditions, the analyst can perform robust and flexible analysis with our methods.
In addition, and this is a result that further highlights our contributions, we show that the influence function in the maximum likelihood method does not have redescendence for any link function in the ordinal response model.
Through numerical experiments using artificial and two real data, we show that the proposed methods perform better than the maximum likelihood method with and without outliers in the data for various link functions.
\end{abstract}

\noindent{{\bf Keywords}: density-power divergence; $\gamma$-divergence; influence function; link function; ordered category; outlier; robust inference}

\medskip

\noindent{{\bf Mathematics Subject Classification}: Primary 62F35 ; Secondary 62J12}
\section{Introduction}
\label{sec:intro}
Ordered categorical data has become popular in a wide range of fields such as medicine, sociology, psychology, political sciences, economics, marketing, and so on (\citealp{breiger1981social, ashby1989ordered, uebersax1999probit}). 
Ordered categorical data are, for example, the progression of a disease expressed as stage 1, 2, 3, or 4, or opinions on a policy expressed as opposition, neutrality, or approval.
In addition, when continuous data are summarized into categorical data, such as ages 0-20, 21-40, 41-60, 61-80, and 80 or more, the categorical data are ordered categorical data.
For this reason, ordered categorical data are often considered to be discretized values of latent continuous variables.

There have been many studies on how to analyze ordered categorical data from half a century ago to the present (\citealp{mccullagh1980regression, albert1993bayesian, tomizawa2006decompositions, agresti2010analysis, agresti2017ordinal, baetschmann2020feologit}).
This paper focuses on one of the most commonly used methods, the ordinal response model, that is a regression on ordered categorical data as a response variable.
The ordinal response model is one of the frameworks of generalized linear models, and is called the ordinal regression model, or the cumulative link model since it connects the cumulative probability of belonging to a certain category to the covariates with the ``link'' function (\citealp{mccullagh1980regression}).
The link functions most often used in data analysis are the probit link, which is the distribution function of the standard normal distribution, the logit link, which is the distribution function of the standard logistic distribution, the log-log link, which is the distribution function of the right-skewed Gumbel distribution, and the complementary log-log link, which is the distribution function of the left-skewed log-Weibull distribution.
The formulation of the ordinal response model and its details are described in Section \ref{sec:orm}.

Outliers can be caused by various reasons, such as typos in the values and misrecognition of units.
Since it is well known that the maximum likelihood method is strongly affected by outliers, another statistical method that can appropriately deal with outliers is required nowadays.
There have been many studies on how to deal with outliers, for example, the well-known inferences based on Huber-type loss and robust divergences (\citealp{hampel1986robust, basu1998robust, jones2001comparison, fujisawa2008robust, huber2009robust, ghosh2013robust, maronna2019robust, castilla2021estimation}), but most of them have focused on continuous, binary, or counted data, with few focusing on ordered categorical data.

Of course, the ordinal response model is not an exception to the model affected by outliers, although the response variable is discrete type data.
The maximum likelihood method in the ordinal response model with the commonly used link functions, such as above mentioned is strongly affected by outliers.
Although outliers in discrete data may be harder to imagine than in continuous data, an "outlier" in the ordered response model is defined as a combination of ordered categorical data and its covariates that is heterogeneous relative to other pairs (\citealp{riani2011outliers}).
This may be taken to mean that the combination of ordered categorical data and its covariates is inconsistent.
Unless otherwise noted, we use the same definition of outliers in this paper.

There are various methods to check whether an inference method is robust against outliers, and often an influence function (\citealp{hampel1974influence}) may be used in linear regression.
Simply put, the influence function is an index to check the ``influence'' of data on an inference method, and its value must be bounded at least to be robust against outliers.
This is because if the value diverges for a given data, the result of the inference depends almost entirely on the data without regard to other data.

As mentioned earlier, there are few studies on outliers in the ordinal response model, but there are studies that derive the conditions for the inference method and link function such that the influence function is bounded in the ordinal response model as well.
\cite{croux2013robust} and \cite{iannario2017robust} proposed the weighted maximum likelihood methods using the Student, 0/1, and Huber weights so that the influence function in the ordinal response model is bounded.
However, these papers do not specify the algorithms to perform their inference methods, which makes it difficult for practitioners who wish to implement the robust ordinal response model.
We also attempted to implement their inference methods, but found it extremely difficult to converge the numerical optimization algorithm to compute their estimators.

\cite{scalera2021robust} derived a class of link functions for bounded influence functions in the maximum likelihood method in the ordinal response model.
However, the class does not include probit, logit, log-log and complementary log-log link functions which are commonly used in the analysis.
That is, analysts cannot perform robust and flexible modeling for ordered categorical data.
We have also confirmed that misspecification of the link function causes a substantial bias in the parameter estimation, although this is not discussed in depth here.
This is also the case in the framework of binary regression, which is discussed in \cite{czado1992effect}, and confirms that the flexibility in the choice of the link function is very important.

The boundedness of the influence function is an important property to ensure that the result of inference is not too much sensitive by only certain data, but it does not mean that data that are largely outliers compared to other data do not influence the result of inference at all.
In other words, even if the influence function is bounded, there is still some influence from outliers.
Therefore, it is desirable for the influence function to satisfy not only boundedness but also redescendence (\citealp{maronna2019robust}), which means that the influence of large outliers on inference can be ignored.
In this paper, we consider a class of link functions for the influence function to satisfy redescendence in the ordinal response model in the maximum likelihood method, and show that such a class does not exist (Theorem \ref{thm:if} in Section \ref{sec:if}).

Since the influence function for the maximum likelihood method in the ordinal response model does not satisfy redescendence, it is necessary to consider another inference method.
The weighted maximum likelihood methods of \cite{croux2013robust} and \cite{iannario2017robust} do not satisfy redescendence because the weights used are the Student, 0/1, and Huber types (\citealp{maronna2019robust}), although the influence function is bounded.
Thus, we propose inference methods in the ordinal response model with two robust divergences, the density-power (DP) (\citealp{basu1998robust}) and $\gamma$-divergences (\citealp{jones2001comparison, fujisawa2008robust}), which are expected to satisfy the redescendence of their influence functions.

Although recently, \cite{pyne2022robust} proposed an inference method in the ordinal response model using the DP divergence, they did not discuss the condition for link functions to satisfy boundedness and redescendence of the influence function in their method.
Our contribution is not only to derive the condition for link functions satisfying boundedness and redescendence of the influence function in the DP divergence (Theorem \ref{thm:dp_if} in Section \ref{sec:if}), but also to propose an inference method in the ordinal response model using the $\gamma$-divergence, which is also commonly used for robust inference, and derive the condition of link functions to achieve robust inference (Theorem \ref{thm:g_if} in Section \ref{sec:if}), and compare the performance between the DP and $\gamma$-divergences.
The derivation of this condition provides a guideline for practitioners to decide which of the various link functions to use when conducting robust analysis against outliers using the ordinal response model.
It will also help to answer the question of whether the DP or $\gamma$-divergences should be used.
Of course, the algorithm of our proposed methods can be used by anyone since the programming code described using the $\mathrm{R}$ is available on the GitHub repository (\url{https://github.com/t-momozaki/RORM}).


The paper is organized as follows
Section \ref{sec:orm} introduces the ordinal response model and its maximum likelihood method.
Section \ref{sec:div} introduces two robust divergences, the DP and $\gamma$-divergences, and proposes estimators based on these divergences.
We also briefly show that the proposed estimators are robust against outliers.
Section \ref{sec:if} derives the influence functions of the proposed methods and considers the robustness of the proposed methods in terms of the influence functions.
Section \ref{sec:asymp} considers the asymptotic properties of the proposed estimators and the testing procedures.
Section \ref{sec:ne} conducts some numerical experiments using artificial and two real data for the proposed methods and evaluates the performance of the proposed methods.
Section \ref{sec:conclude} provides the conclusion and some remarks.


%

\section{Ordinal response model and its maximum likelihood method} 
\label{sec:orm}
In this section, we discuss the ordinal response model and the maximum likelihood method, a parameter estimation method commonly used in the model.

Consider the following latent variable model for $M$ ordered categorical data $y_i$ ($i=1,2,\ldots,n$) as the response variable.
\begin{equation}
\label{eq:lvm}
z_i = \bm{x}_i^\top \bm{\beta} + \ep_i,
\end{equation}
for $i=1,2,\ldots,n$, where $z_i$ is called the (continuous) latent variable, 
\begin{equation*}
y_i = 
\begin{cases}
1 & (\delta_0 < z_i \leq \delta_1) \\
2 & (\delta_1 < z_i \leq \delta_2) \\
\vdots \\
M & (\delta_{M-1} < z_i \leq \delta_M)
\end{cases},
\end{equation*}
$\bm{x}_i = (x_{i1}, x_{i2}, \ldots, x_{ip})^\top$, $\bm{\beta} = (\beta_1, \beta_2, \ldots, \beta_p)^\top$, $\ep_i \overset{i.i.d.}{\sim} G(\cdot)$ (known) and has the density function $g(\cdot)$, and $-\infty=\delta_0<\delta_1<\cdots<\delta_M=\infty$ are the cutpoints.
For the identification, in the $G(\cdot)$ there is no error scale and the latent model has no the intercept term.
Note that the observed data is $(\bm{y}, \bm{X})$, where $\bm{y}=(y_1,y_2,\ldots,y_n)^\top$ and $\bm{X}=(\bm{x}_1,\bm{x}_2,\ldots,\bm{x}_n)^\top$, and the latent variable $\bm{z}=(z_1,z_2,\ldots,z_p)^\top$ is unobserved.
The parameters to be estimated in this model are $\bm{\theta}=(\bm{\beta},\bm{\delta})$, where $\bm{\delta}=(\delta_1,\delta_2,\ldots,\delta_{M-1})$.

The probability mass function of $y_i$ given $\bm{x}_i$ with $\bm{\theta}$ is
\begin{equation*}
f(y_i|\bm{x}_i;\bm{\theta}) = \prod_{m=1}^M \Pr(y_i=m|\bm{x}_i;\bm{\theta})^{I(y_i=m)},
\end{equation*}
where the indicator function $I(\cdot)$ and
\begin{align*}
\Pr(y_i=m|\bm{x}_i;\bm{\theta}) &= \Pr(\delta_{m-1}<z_i\leq\delta_m|\bm{x}_i;\bm{\theta}) \\
&= G(\delta_m-\bm{x}_i^\top\bm{\beta}) - G(\delta_{m-1}-\bm{x}_i^\top\bm{\beta}).
\end{align*}
Namely, the likelihood function for the ordinal response model is expressed as
\begin{equation}
\label{eq:orm}
f(\bm{y}|\bm{X};\bm{\theta}) = \prod_{i=1}^n \left[ G(\delta_{y_i}-\bm{x}_i^\top\bm{\beta}) - G(\delta_{y_i-1}-\bm{x}_i^\top\bm{\beta}) \right].
\end{equation}
The maximum likelihood estimator (MLE) of $\bm{\theta}$ are obtained as $\bm{\theta}$ that minimizes the negative log-likelihood function.
That is, noting the constraint $\delta_1<\delta_2<\cdots<\delta_{M-1}$ the MLE is obtained by
\begin{align*}
\hat{\bm{\theta}}_{ML} &= \argmin_{\bm{\theta}} \left\{ - \log f(\bm{y}|\bm{X};\bm{\theta}) \right\} \\
&= \argmin_{\bm{\theta}} \left\{ - \sum_{i=1}^n \log \left[ G(\delta_{y_i}-\bm{x}_i^\top\bm{\beta}) - G(\delta_{y_i-1}-\bm{x}_i^\top\bm{\beta}) \right] \right\}.
\end{align*}
It is well known that the MLE $\hat{\bm{\theta}}_{ML}$ is strongly affected by the outliers in the observations ($\bm{y},\bm{X}$).
This is because if $|\bm{x}_1^\top\bm{\beta}|$ is large enough, i.e., if $(y_1,\bm{x}_1)$ is outliers compared to the other data, then $\log \left[ G(\delta_{y_1}-\bm{x}_1^\top\bm{\beta}) - G(\delta_{y_1-1}-\bm{x}_1^\top\bm{\beta}) \right]$  takes a large value and the entire objective function is strongly affected by the outlier $(y_1,\bm{x}_1)$.

\begin{rem}[Optimization with the ordering constraint]
There exist various methods for optimization algorithms with the ordering constraint such as the cutpoints in the ordinal response model, including those using matrix notation (\citealp{christensen2018cumulative}), but in this paper we use the \cite{franses2001quantitative}'s method, which reparameterize the cutpoints as follows.
\begin{equation*}
\delta_1 = \tilde{\delta}_1, ~~, \delta_m = \tilde{\delta}_1 + \sum_{j=2}^m \tilde{\delta}_j^2 ~~ \mbox{for $m=2,3,\ldots,M-1$}.
\end{equation*}
Then the likelihood function \eqref{eq:orm} for the ordinal response model can be expressed as
\begin{equation*}
f(\bm{y}|\bm{X};\tilde{\bm{\theta}}) = \prod_{i=1}^n \left[ G\left( \tilde{\delta}_1 + \sum_{j=2}^{y_i} \tilde{\delta}_j^2 -\bm{x}_i^\top\bm{\beta}\right) - G\left( \tilde{\delta}_1 + \sum_{j=2}^{y_i} \tilde{\delta}_j^2 -\bm{x}_i^\top\bm{\beta}\right) \right],
\end{equation*}
where $\tilde{\bm{\theta}} = (\bm{\beta},\tilde{\bm{\delta}})$ with $\tilde{\bm{\delta}}=(\tilde{\delta}_1,\tilde{\delta}_2,\ldots,\tilde{\delta}_{M-1})$, and we can obtain the MLE of $\tilde{\bm{\theta}}$ by
\begin{align*}
\argmin_{\tilde{\bm{\theta}}} \left\{ - \log f(\bm{y}|\bm{X};\tilde{\bm{\theta}}) \right\} .
\end{align*}
This minimization problem can be easily solved by using the \texttt{optim} function in the {\bf stats} library of the {\bf R} programming language.
All the following minimization problems will be solved using the above reparameterization, but to simplify the notation, we will use $\bm{\delta}$ before reparameterization throughout this paper.
\end{rem}

\section{Estimators via robust divergence for the ordinal response model} 
\label{sec:div}
This section introduces two robust divergences, the DP and $\gamma$-divergences, proposes robust estimators for the ordinal response model using these divergences, and describes their properties against outliers.
Section \ref{sec:density} describes the case using the DP divergence, and Section \ref{sec:gamma} describes the case using the $\gamma$-divergence.

\subsection{Case: Density-Power divergence} 
\label{sec:density}
Suppose that $h(x,y)$, $h(y|x)$, and $h(x)$ are the underlying probability density (or mass) functions of $(x,y)$, $y$ given $x$, and $x$, respectively.
Let $f(y|x;\theta)$ be the probability density (or mass) function of $y$ given $x$ with parameter $\theta$.
This paper uses the definition of the DP cross entropy
\begin{align}
&d_{DP}(h(y|x), f(y|x;\theta); h(x)) \nonumber \\
&= -\frac{1}{\alpha} \int \int f(y|x;\theta)^{\alpha} h(x,y) dx dy + \frac{1}{1+\alpha} \int \left( \int f(y|x;\theta)^{1+\alpha} dy \right) h(x) dx \label{eq:dpce}
\end{align}
for $\alpha > 0$.
Using the DP cross entropy, the DP divergence is defined as 
\begin{equation*}
D_{DP}(h(y|x), f(y|x;\theta); h(x)) = -d_{DP}(h(y|x), h(y|x); h(x)) + d_{DP}(h(y|x), f(y|x;\theta); h(x)).
\end{equation*}
About the properties of the DP divergence, see \cite{basu1998robust}.

Using the DP divergence, the target parameter can be considered by
\begin{align*}
\theta_{DP}^* &= \argmin_{\theta} D_{DP}(h(y|x), f(y|x;\theta); h(x)) \\
&= \argmin_{\theta} d_{DP}(h(y|x), f(y|x;\theta); h(x)),
\end{align*}
and when $h(y|x)=f(y|x;\theta^*)$, $\theta_{DP}^*=\theta^*$.

Suppose that the observed data $(x_1,y_1),(x_2,y_2),\ldots,(x_n,y_n)$ are randomly drawn from the underlying distribution $h(x,y)$.
Then, from the formulation \eqref{eq:dpce}, the DP cross entropy is empirically estimated by
\begin{equation*}
\tilde{d}_{DP}(f(y|x;\theta)) = -\frac{1}{\alpha} \left\{ \frac{1}{n} \sum_{i=1}^n f(y_i|x_i;\theta)^{\alpha} \right\} + \frac{1}{1+\alpha} \left\{ \frac{1}{n} \sum_{i=1}^n \int  f(y|x_i;\theta)^{1+\alpha} dy \right\}.
\end{equation*}
Namely, the DP estimator is defined by
\begin{equation*}
\hat{\theta}_{DP} = \argmin_{\theta} \tilde{d}_{DP}(f(y|x;\theta)).
\end{equation*}

Based on the above, we propose the following DP-estimator for the ordinal response model.
\begin{equation}
\label{eq:dp_est}
\hat{\bm{\theta}}_{DP} = \argmin_{\bm{\theta}} \tilde{d}_{DP}(f(\bm{y}|\bm{X};\bm{\theta})),
\end{equation}
where the empirically estimated DP cross entropy
\begin{equation}
\label{eq:dpce_or}
\begin{split}
\tilde{d}_{DP}(f(\bm{y}|\bm{X};\bm{\theta})) =& -\frac{1}{\alpha} \left\{ \frac{1}{n} \sum_{i=1}^n \left[ G(\delta_{y_i}-\bm{x}_i^\top\bm{\beta}) - G(\delta_{y_i-1}-\bm{x}_i^\top\bm{\beta}) \right]^{\alpha} \right\} \\
&+ \frac{1}{1+\alpha} \left\{ \frac{1}{n} \sum_{i=1}^n \sum_{m=1}^M  \left[ G(\delta_m-\bm{x}_i^\top\bm{\beta}) - G(\delta_{m-1}-\bm{x}_i^\top\bm{\beta}) \right]^{1+\alpha} \right\}.
\end{split}
\end{equation}

We briefly confirm that the proposed DP-estimator is robust against outliers.
Suppose that $|\bm{x}_1^\top\bm{\beta}|$ is large enough, i.e., $(y_1,\bm{x}_1)$ is outliers compared to the other data.
Then, the empirically estimated DP cross entropy \eqref{eq:dpce_or} is expressed as
\begin{align*}
\tilde{d}_{DP}(f(\bm{y}|\bm{X};\bm{\theta})) \approx& -\frac{1}{\alpha} \left\{ \frac{1}{n-1} \sum_{i=2}^n \left[ G(\delta_{y_i}-\bm{x}_i^\top\bm{\beta}) - G(\delta_{y_i-1}-\bm{x}_i^\top\bm{\beta}) \right]^{\alpha} \right\} \\
&+ \frac{1}{1+\alpha} \left\{ \frac{1}{n-1} \sum_{i=2}^n \sum_{m=1}^M  \left[ G(\delta_m-\bm{x}_i^\top\bm{\beta}) - G(\delta_{m-1}-\bm{x}_i^\top\bm{\beta}) \right]^{1+\alpha} \right\}
\end{align*}
since $G(\delta_{y_1}-\bm{x}_1^\top\bm{\beta}) - G(\delta_{y_1-1}-\bm{x}_1^\top\bm{\beta})$ takes a small value.
Namely, the objective function $\tilde{d}_{DP}(f(\bm{y}|\bm{X};\bm{\theta}))$ in \eqref{eq:dp_est} is expressed by removing the outliers $(y_1,\bm{x}_1)$, and as a result, the effect of the outliers can be ignored in the parameter estimation.


\subsection{Case: $\gamma$-divergence} 
\label{sec:gamma}
Suppose that $h(x,y)$, $h(y|x)$, and $h(x)$ are the underlying probability density (or mass) functions of $(x,y)$, $y$ given $x$, and $x$, respectively.
Let $f(y|x;\theta)$ be the probability density (or mass) function of $y$ given $x$ with parameter $\theta$.
This paper uses the definition of the $\gamma$-cross entropy of \cite{kawashima2017robust} as follows.
\begin{align}
&d_{\gamma}(h(y|x), f(y|x;\theta); h(x)) \nonumber \\
&= -\frac{1}{\gamma} \log \int \left( \int h(y|x) f(y|x;\theta)^{\gamma} dy \right) h(x) dx + \frac{1}{1+\gamma} \log \int \left( \int f(y|x;\theta)^{1+\gamma} dy \right) h(x) dx \nonumber \\
&= -\frac{1}{\gamma} \log \int \int f(y|x;\theta)^{\gamma} h(x,y) dx dy + \frac{1}{1+\gamma} \log \int \left( \int f(y|x;\theta)^{1+\gamma} dy \right) h(x) dx \label{eq:gce}
\end{align}
for $\gamma>0$.
Using the $\gamma$-cross entropy, the $\gamma$-divergence is defined as follows.
\begin{equation*}
D_{\gamma}(h(y|x), f(y|x;\theta); h(x)) = -d_{\gamma}(h(y|x), h(y|x); h(x)) + d_{\gamma}(h(y|x), f(y|x;\theta); h(x)).
\end{equation*}
About the properties of the $\gamma$-divergence, see \cite{fujisawa2008robust} and \cite{kawashima2017robust}.

Using the $\gamma$-divergence, the target parameter can be considered by
\begin{align*}
\theta_{\gamma}^* &= \argmin_{\theta} D_{\gamma}(h(y|x), f(y|x;\theta); h(x)) \\
&= \argmin_{\theta} d_{\gamma}(h(y|x), f(y|x;\theta); h(x)),
\end{align*}
and when $h(y|x)=f(y|x;\theta^*)$, $\theta_{\gamma}^*=\theta^*$.

Suppose that the observed data $(x_1,y_1),(x_2,y_2),\ldots,(x_n,y_n)$ are randomly drawn from the underlying distribution $h(x,y)$.
Then, from the formulation \eqref{eq:gce}, the $\gamma$-cross entropy is empirically estimated by
\begin{equation*}
\tilde{d}_{\gamma}(f(y|x;\theta)) = -\frac{1}{\gamma} \log \left\{ \frac{1}{n} \sum_{i=1}^n f(y_i|x_i;\theta)^{\gamma} \right\} + \frac{1}{1+\gamma} \log \left\{ \frac{1}{n} \sum_{i=1}^n \int  f(y|x_i;\theta)^{1+\gamma} dy \right\}.
\end{equation*}
Namely, the $\gamma$-estimator is defined by
\begin{equation}
\label{eq:g_est}
\hat{\theta}_{\gamma} = \argmin_{\theta} \tilde{d}_{\gamma}(f(y|x;\theta)).
\end{equation}

Based on the above, we propose the following $\gamma$-estimator for the ordinal response model.
\begin{equation}
\label{eq:gest_or}
\hat{\bm{\theta}}_{\gamma} = \argmin_{\bm{\theta}} \tilde{d}_{\gamma}(f(\bm{y}|\bm{X};\bm{\theta})),
\end{equation}
where the empirically estimated $\gamma$-cross entropy
\begin{equation}
\label{eq:gce_or}
\begin{split}
\tilde{d}_{\gamma}(f(\bm{y}|\bm{X};\bm{\theta})) =& -\frac{1}{\gamma} \log \left\{ \frac{1}{n} \sum_{i=1}^n \left[ G(\delta_{y_i}-\bm{x}_i^\top\bm{\beta}) - G(\delta_{y_i-1}-\bm{x}_i^\top\bm{\beta}) \right]^{\gamma} \right\} \\
&+ \frac{1}{1+\gamma} \log \left\{ \frac{1}{n} \sum_{i=1}^n \sum_{m=1}^M  \left[ G(\delta_m-\bm{x}_i^\top\bm{\beta}) - G(\delta_{m-1}-\bm{x}_i^\top\bm{\beta}) \right]^{1+\gamma} \right\}.
\end{split}
\end{equation}

We briefly confirm that the proposed $\gamma$-estimator is robust against outliers.
Suppose that $|\bm{x}_1^\top\bm{\beta}|$ is large enough, i.e., $(y_1,\bm{x}_1)$ is outliers compared to the other data.
Then, the empirically estimated $\gamma$-cross entropy \eqref{eq:gce_or} is expressed as
\begin{align*}
\tilde{d}_{\gamma}(f(\bm{y}|\bm{X};\bm{\theta})) \approx& -\frac{1}{\gamma} \log \left\{ \frac{1}{n-1} \sum_{i=2}^n \left[ G(\delta_{y_i}-\bm{x}_i^\top\bm{\beta}) - G(\delta_{y_i-1}-\bm{x}_i^\top\bm{\beta}) \right]^{\gamma} \right\} \\
&+ \frac{1}{1+\gamma} \log \left\{ \frac{1}{n-1} \sum_{i=2}^n \sum_{m=1}^M  \left[ G(\delta_m-\bm{x}_i^\top\bm{\beta}) - G(\delta_{m-1}-\bm{x}_i^\top\bm{\beta}) \right]^{1+\gamma} \right\}
\end{align*}
since $G(\delta_{y_1}-\bm{x}_1^\top\bm{\beta}) - G(\delta_{y_1-1}-\bm{x}_1^\top\bm{\beta})$ takes a small value.
Namely, the objective function $\tilde{d}_{\gamma}(f(\bm{y}|\bm{X};\bm{\theta}))$ in \eqref{eq:gest_or} is expressed by removing the outliers $(y_1,\bm{x}_1)$, and as a result, the effect of the outliers can be ignored in the parameter estimation.


\section{Influence functions for the ordinal response model}
\label{sec:if}
The influence function is classically often used as a measure of robustness of an estimator, and it is required that the influence function is bounded for the observations (see, \citealp{hampel1974influence}).
This is because the influence function represents the effect on the estimator when there are a few outliers in the observed values.
Furthermore, it preferably has the redescendence, that is the property that large outliers have little effect on the parameter estimation (\citealp{maronna2019robust}).

This section derives the respective influence functions in the ordinal response model using the DP and $\gamma$-divergences, and discusses the robustness of each proposed estimation method against outliers in terms of the influence functions.
The influence functions are plotted and compared.
Before that, we describe notations for the discussion and discuss the influence function in the maximum likelihood method.

Suppose that $(y_1,\bm{x}_1),(y_2,\bm{x}_2),\ldots,(y_n,\bm{x}_n)$ are generated from the true distribution $F(y,\bm{x})$, and consider the contaminated model $F_{\rho}(y,\bm{x}) = (1-\rho)F(y,\bm{x}) + \rho \Delta_{(y_o,\bm{x}_o)}(y,\bm{x})$, where $\rho$ is the contamination ratio and $\Delta_{(y_o,\bm{x}_o)}(y,\bm{x})$ is the contaminating distribution degenerated at $(y_o,\bm{x}_o)$.
The influence function in the ordinal response model using the maximum likelihood method is expressed as follows.
\begin{equation*}
IF_{ML}(y_o,\bm{x}_o; F, \bm{\theta}) = \left( \bm{M}_{ML}(\bm{\theta},\psi_{ML}) \right)^{-1} \psi_{ML}(y_o|\bm{x}_o;\bm{\theta}),
\end{equation*}
where
\begin{equation*}
\psi_{ML}(y_o|\bm{x}_o;\bm{\theta}) = s(y_o|\bm{x}_o;\bm{\theta}) = \frac{\pd \log f(y_o|\bm{x}_o;\bm{\theta})}{\pd \bm{\theta}},
\end{equation*}
\begin{equation*}
\bm{M}_{ML}(\bm{\theta},\psi_{ML}) = - n \E_{(y,\bm{x})} \left[ \frac{\pd}{\pd \bm{\theta}} \psi_{ML}(y|\bm{x};\bm{\theta}) \right] = n \mathcal{I}(\bm{\theta}),
\end{equation*}
and $\mathcal{I}(\bm{\theta})$ is the fisher information matrix.
Note that since $\E_{(y,\bm{x})}$ is the expected value for the distribution $F(y,\bm{x})$, $\bm{M}_{ML}(\bm{\theta},\psi_{ML})$ is constant and bounded with respect to $(y,\bm{x})$.
That is, for the boundedness of the influence function $IF_{ML}(y_o,\bm{x}_o; F, \bm{\theta})$, the function $\psi_{ML}(y_o|\bm{x}_o;\bm{\theta})$ must be bounded.

The influence function of the parameter $\beta_k$ ($k=1,2,\ldots,p$) in the maximum likelihood method for the ordinal response model is expressed as follows.
\begin{equation}
\label{eq:ml_b_if}
IF_{ML}(y_o,\bm{x}_o; F, \beta_k) \propto \psi_{ML}(y_o|\bm{x}_o;\beta_k) = -\frac{ g(\delta_{y_o}-\bm{x}_o^\top\bm{\beta}) - g(\delta_{y_o-1}-\bm{x}_o^\top\bm{\beta}) }{ G(\delta_{y_o}-\bm{x}_o^\top\bm{\beta}) - G(\delta_{y_o-1}-\bm{x}_o^\top\bm{\beta}) } x_{ok}.
\end{equation}
The influence function of the parameter $\delta_l$ ($l=1,2,\ldots,M-1$) in the maximum likelihood method for the ordinal response model is expressed as follows.
\begin{equation}
\label{eq:ml_d_if}
IF_{ML}(y_o,\bm{x}_o; F, \delta_l) \propto \psi_{ML}(y_o|\bm{x}_o;\delta_l) = \frac{ g(\delta_{l}-\bm{x}_o^\top\bm{\beta}) \left[ I(y_o=l) - I(y_o=l+1) \right] }{ G(\delta_{y_o}-\bm{x}_o^\top\bm{\beta}) - G(\delta_{y_o-1}-\bm{x}_o^\top\bm{\beta}) }.
\end{equation}

In order for these influence functions \eqref{eq:ml_b_if} and \eqref{eq:ml_d_if} to be bounded in the maximum likelihood method, the conditions for the distribution $G$ of $\ep_i$ in the latent variable model \eqref{eq:lvm} were derived by \cite{scalera2021robust}.
\begin{thm}[Proposition 3.2 of \citealp{scalera2021robust}]
The influence function for $\bm{\beta}$ \eqref{eq:ml_b_if} is bounded if and only if
\begin{equation}
\label{cod:bn_ml_b}
\lim_{u\to\pm\infty} \left| u \frac{\pd \log g(u)}{\pd u} \right| < +\infty,
\end{equation}
where the $g(\cdot)$ is the density function of $G$.
\end{thm}

\begin{thm}[Proposition 3.1 of \citealp{scalera2021robust}]
The influence function for $\bm{\delta}$ \eqref{eq:ml_d_if} is bounded if and only if
\begin{equation}
\label{cod:bn_ml_d}
\lim_{u\to\pm\infty} \left| \frac{\pd \log g(u)}{\pd u} \right| < +\infty,
\end{equation}
where the $g(\cdot)$ is the density function of $G$.
\end{thm}

\noindent
The distributions $G$ that satisfy the conditions \eqref{cod:bn_ml_b} and \eqref{cod:bn_ml_d} include the Student's $t$-distribution where the degree of freedom $\nu$ is sufficiently small (\citealp{scalera2021robust}).
When the distribution $G$ is the Student's $t$-distribution, since the influence function of $\beta_k$ is $-\nu-1$ for large outliers, the influence function of $\beta_k$ is bounded for $\nu<+\infty$, however the parameter estimation is always affected by outliers constantly.
Of course, the same is true for $\delta_l$.

Although \cite{scalera2021robust} mention the boundedness of the influence functions in the maximum likelihood method, they do not mention redescendence.
Hence, we further give the following theorem on the redescendence of the influence functions in the maximum likelihood method.
\begin{thm}
\label{thm:if}
The influence functions for $\bm{\beta}$ \eqref{eq:ml_b_if} and $\bm{\delta}$ \eqref{eq:ml_d_if} in the maximum likelihood method for the ordinal response model are not redescendence.
\end{thm}
\noindent
The proof is given in Appendix \ref{app:ml_if}.

\begin{rem}[Redescendence of the influence function of the maximum likelihood method in the ordinal response model]
There are various studies on heavy-tailed distributions, and recently the log-regularly varying function (\citealp{desgagne2015robustness}) whose order of tail is same as $(u \log u)^{-1}$, such as the log-Pareto truncated normal distribution proposed by \cite{gagnon2020new} and the extremely heavily-tailed distribution proposed by \cite{hamura2022log} is well known. 
However, even if these distributions are applied to $G(\cdot)$, the influence function of $\beta_k$ for the maximum likelihood method in the ordinal response model do not have redescendence.
It is also known that $g(\delta_{l}-\bm{x}_i^\top\bm{\beta})/g(\delta^*-\bm{x}_i^\top\bm{\beta})$ is of constant order for regularly varying and log-regularly varying functions (\citealp{o2012bayesian, desgagne2015robustness}).
Therefore, when Student's $t$-distribution or the heavy tailed distributions described above is applied to the distribution $G(\cdot)$, when $y_i=1$ and $l=1$, or $y_i=M$ and $l=M-1$, the influence function of $\delta_l$ is zero for large outliers.
\end{rem}

\subsection{Case: Density-Power divergence}
\label{sec:dp_if}
In the similar manner as in the maximum likelihood method, the influence function using the DP divergence is expressed as follows.
\begin{equation*}
IF_{DP}(y_o,\bm{x}_o; F, \bm{\theta}) \propto \psi_{DP}(y_o|\bm{x}_o;\bm{\theta}),
\end{equation*}
where
\begin{equation*}
\psi_{DP}(y_o|\bm{x}_o;\bm{\theta}) = f(y_o|\bm{x}_o;\bm{\theta})^{\alpha} s(y_o|\bm{x}_o;\bm{\theta}) - \E_{(y|\bm{x})}[ f(y|\bm{x};\bm{\theta})^{\alpha} s(y|\bm{x};\bm{\theta}) ]
\end{equation*}
and $\E_{(y|\bm{x})}$ is the conditional expectation for the conditional distribution $F(y|\bm{x})$.
Thus, the influence functions of the parameters $\beta_k$ ($k=1,2,\ldots,p$) and $\delta_l$ ($l=1,2,\ldots,M-1$) for the ordinal response model with the DP divergence are expressed by
\begin{equation}
\label{eq:dp_b_if}
\begin{split}
&IF_{DP}(y_o,\bm{x}_o; F, \beta_k) \propto \psi_{DP}(y_o|\bm{x}_o;\beta_k) \\
=& - \left[ g(\delta_{y_o}-\bm{x}_o^\top\bm{\beta}) - g(\delta_{y_o-1}-\bm{x}_o^\top\bm{\beta}) \right] \left[ G(\delta_{y_o}-\bm{x}_o^\top\bm{\beta}) - G(\delta_{y_o-1}-\bm{x}_o^\top\bm{\beta}) \right]^{\alpha-1} x_{ok} \\
&+ \left( \sum_{m=1}^M \left[ g(\delta_{m}-\bm{x}_o^\top\bm{\beta}) - g(\delta_{m-1}-\bm{x}_o^\top\bm{\beta}) \right] \left[ G(\delta_{m}-\bm{x}_o^\top\bm{\beta}) - G(\delta_{m-1}-\bm{x}_o^\top\bm{\beta}) \right]^{\alpha} \right) x_{ok}
\end{split}
\end{equation}
and  
\begin{equation}
\label{eq:dp_d_if}
\begin{split}
&IF_{DP}(y_o,\bm{x}_o; F, \delta_l) \propto \psi_{DP}(y_o|\bm{x}_o;\delta_l) \\
=& g(\delta_{l}-\bm{x}_o^\top\bm{\beta}) \left[ G(\delta_{y_o}-\bm{x}_o^\top\bm{\beta}) - G(\delta_{y_o-1}-\bm{x}_o^\top\bm{\beta}) \right]^{\alpha-1} \left[ I(y_o=l) - I(y_o=l+1) \right] \\
&- g(\delta_{l}-\bm{x}_o^\top\bm{\beta}) \\
&\quad \times \left( \left[ G(\delta_{l}-\bm{x}_o^\top\bm{\beta}) - G(\delta_{l-1}-\bm{x}_o^\top\bm{\beta}) \right]^{\alpha} - \left[ G(\delta_{l+1}-\bm{x}_o^\top\bm{\beta}) - G(\delta_{l}-\bm{x}_o^\top\bm{\beta}) \right]^{\alpha} \right),
\end{split}
\end{equation}
respectively.

Here, the following theorem holds for the influence functions \eqref{eq:dp_b_if} and \eqref{eq:dp_d_if} using the DP divergence.
\begin{thm}
\label{thm:dp_if}
If there exists a certain $0<\alpha\leq1$ such that 
\begin{equation}
\label{eq:cod_dp_if}
\lim_{u\to\pm\infty} g(u)^\alpha u = 0,
\end{equation}
then the influence functions \eqref{eq:dp_b_if} and \eqref{eq:dp_d_if} are bounded and redescendent.
\end{thm}
\noindent
The proof is given in Appendix \ref{app:dp_if}.

Theorem \ref{thm:dp_if} ensures that the proposed method using the DP divergence is robust against outliers in terms of the influence function.
That is, the estimator of the proposed method is hardly affected by the large outliers.

\subsection{Case: $\gamma$-divergence}
\label{sec:g_if}
In the similar manner as in the maximum likelihood method, the influence function using the $\gamma$-divergence is expressed as follows.
\begin{equation*}
IF_{\gamma}(y_o,\bm{x}_o; F, \bm{\theta}) \propto \psi_{\gamma}(y_o|\bm{x}_o;\bm{\theta}),
\end{equation*}
where
\begin{align*}
\psi_{\gamma}(y_o|\bm{x}_o;\bm{\theta}) = f(y_o|\bm{x}_o;\bm{\theta})^{\gamma} s(y_o|\bm{x}_o;\bm{\theta}) \E_{(y|\bm{x})}[ f(y|\bm{x};\bm{\theta})^{\gamma} ] - f(y_o|\bm{x}_o;\bm{\theta})^{\gamma} \E_{(y|\bm{x})}[ f(y|\bm{x};\bm{\theta})^{\gamma} s(y|\bm{x};\bm{\theta}) ].
\end{align*}
Thus, the influence functions of the parameters $\beta_k$ ($k=1,2,\ldots,p$) and $\delta_l$ ($l=1,2,\ldots,M-1$) for the ordinal response model with the $\gamma$-divergence are expressed by
\begin{equation}
\label{eq:g_b_if}
\begin{split}
&IF_{\gamma}(y_o,\bm{x}_o; F, \beta_k) \propto \psi_{\gamma}(y_o|\bm{x}_o;\beta_k) \\
=& - \left( \sum_{m=1}^M  \left[ G(\delta_{m}-\bm{x}_o^\top\bm{\beta}) - G(\delta_{m-1}-\bm{x}_o^\top\bm{\beta}) \right]^{\gamma+1} \right) \\
&\quad \times \left[ g(\delta_{y_o}-\bm{x}_o^\top\bm{\beta}) - g(\delta_{y_o-1}-\bm{x}_o^\top\bm{\beta}) \right] \left[ G(\delta_{y_o}-\bm{x}_o^\top\bm{\beta}) - G(\delta_{y_o-1}-\bm{x}_o^\top\bm{\beta}) \right]^{\gamma-1} x_{ok} \\
&+ \left[ G(\delta_{y_o}-\bm{x}_o^\top\bm{\beta}) - G(\delta_{y_o-1}-\bm{x}_o^\top\bm{\beta}) \right]^{\gamma} x_{ok} \\
&\quad \times \left( \sum_{m=1}^M \left[ g(\delta_{m}-\bm{x}_o^\top\bm{\beta}) - g(\delta_{m-1}-\bm{x}_o^\top\bm{\beta}) \right] \left[ G(\delta_{m}-\bm{x}_o^\top\bm{\beta}) - G(\delta_{m-1}-\bm{x}_o^\top\bm{\beta}) \right]^{\gamma} \right)
\end{split}
\end{equation}
and  
\begin{equation}
\label{eq:g_d_if}
\begin{split}
&IF_{\gamma}(y_o,\bm{x}_o; F, \delta_l) \propto \psi_{\gamma}(y_o|\bm{x}_o;\delta_l) \\
=& \left( \sum_{m=1}^M  \left[ G(\delta_{m}-\bm{x}_o^\top\bm{\beta}) - G(\delta_{m-1}-\bm{x}_o^\top\bm{\beta}) \right]^{\gamma+1} \right) \\
&\quad \times g(\delta_{l}-\bm{x}_o^\top\bm{\beta}) \left[ G(\delta_{y_o}-\bm{x}_o^\top\bm{\beta}) - G(\delta_{y_o-1}-\bm{x}_o^\top\bm{\beta}) \right]^{\gamma-1} \left[ I(y_o=l) - I(y_o=l+1) \right] \\
&- g(\delta_{l}-\bm{x}_o^\top\bm{\beta}) \left[ G(\delta_{y_o}-\bm{x}_o^\top\bm{\beta}) - G(\delta_{y_o-1}-\bm{x}_o^\top\bm{\beta}) \right]^{\gamma} \\
&\quad \times \left( \left[ G(\delta_{l}-\bm{x}_o^\top\bm{\beta}) - G(\delta_{l-1}-\bm{x}_o^\top\bm{\beta}) \right]^{\gamma} - \left[ G(\delta_{l+1}-\bm{x}_o^\top\bm{\beta}) - G(\delta_{l}-\bm{x}_o^\top\bm{\beta}) \right]^{\gamma} \right),
\end{split}
\end{equation}
respectively.

Here, the following theorem holds for the influence functions \eqref{eq:g_b_if} and \eqref{eq:g_d_if} using the $\gamma$-divergence.
\begin{thm}
\label{thm:g_if}
If there exists a certain $0<\gamma\leq1$ such that 
\begin{equation}
\label{eq:cod_g_if}
\lim_{u\to\pm\infty} g(u)^\gamma u = 0,
\end{equation}
then the influence functions \eqref{eq:g_b_if} and \eqref{eq:g_d_if} are bounded and redescendent.
\end{thm}
\noindent
The proof of Theorem \ref{thm:g_if} is omitted since it can be done in the same way as the proof of Theorem \ref{thm:dp_if}.

Theorem \ref{thm:g_if} ensures that the proposed method using the $\gamma$-divergence is robust against outliers in terms of the influence function.
That is, the estimator of the proposed method is hardly affected by the large outliers.

\subsection{Numerical experience of influence functions}
Sections \ref{sec:dp_if} and \ref{sec:g_if} prove that the influence functions for the ordinal response model using the DP and $\gamma$-divergences are bounded and redescendent under the conditions \eqref{eq:cod_dp_if} and \eqref{eq:cod_g_if}.
In this section, we plot and compare these influence functions.
Note that when $\alpha=0$ and $\gamma=0$, the methods with the DP and $\gamma$-divergences are the equivalent to the maximum likelihood method, respectively.

The following settings are considered in plotting the influence functions.
The ordinal response $y$ generated by \eqref{eq:lvm}, where $\beta=1.0$ (i.e., using one covariate) and $\bm{\delta}=(-1.5,0.5,1.5)$, has 4 categories.
For the distribution $G$ of the error term in \eqref{eq:lvm}, the standard normal distribution and the standard logistic distribution are used.

\begin{figure}[H]
\includegraphics[scale=0.45]{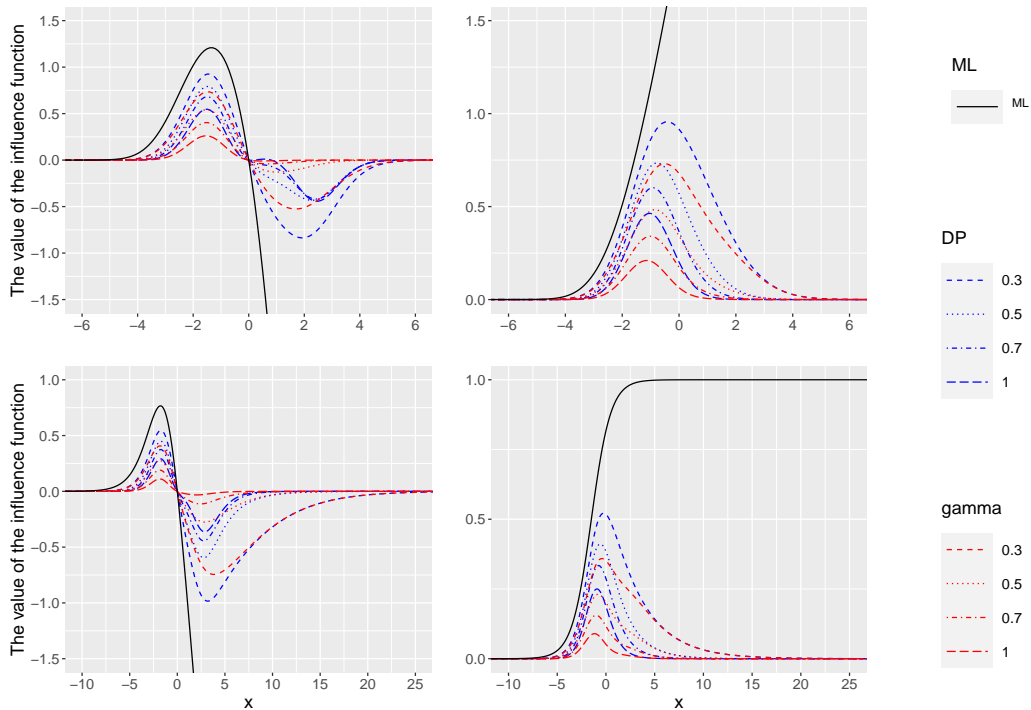}
\caption{Plot of influence function for each parameter with the maximum likelihood (black), the DP divergence (blue) and the $\gamma$-divergence (red): the horizontal is the value of covariates, the vertical is the value of influence function: the first row for the distribution $G$ with standard normal distribution, the second row for standard logistic distribution: the first column is for the coefficient parameter $\beta$, the second column is for the cutoff parameter $\delta_1$}
\label{fg:plot_if}
\end{figure}

Figure \ref{fg:plot_if} plots the influence function (vertical axis) for each parameter against the covariate $x$ (horizontal axis) with the maximum likelihood (black), the DP divergence (blue), and the $\gamma$-divergence (red).
The first and second rows of Figure \ref{fg:plot_if} show the standard normal and standard logistic distributions for $G$, respectively.
The first and second columns are the influence function of $\beta$ and $\delta_1$, respectively.
The line type differs according to the tuning parameter in the divergences, “dash" for $\alpha=\gamma=0.3$, “dot" for $\alpha=\gamma=0.5$, “dot-dash" when $\alpha=\gamma=0.7$, and “long-dash" when $\alpha=\gamma=1.0$.
In plotting the influence functions, we set $y=1$.

Since conditions \eqref{eq:cod_dp_if} and \eqref{eq:cod_g_if} are satisfied when distribution $G$ is the standard normal distribution or the standard logistic distribution, as can be seen from Figure \ref{fg:plot_if}, the influence functions with the DP and  $\gamma$-divergences are both bounded and redescendence.
On the other hand, since the standard normal distribution does not satisfy the conditions \eqref{cod:bn_ml_b} and \eqref{cod:bn_ml_d}, the influence function in the maximum likelihood method is not bounded, as can be seen from Figure \ref{fg:plot_if}, and since the standard logistic distribution satisfies only the condition \eqref{cod:bn_ml_d}, the influence function of $\beta$ is not bounded and that of $\delta_1$ is bounded.

In addition, since the order of $g(u)^\alpha u$ (or $g(u)^\gamma u$) is smaller when the distribution $G$ is lighter in conditions \eqref{eq:cod_dp_if} and \eqref{eq:cod_g_if}, from the proof of Theorems \ref{thm:dp_if} and \ref{thm:g_if} (see Appendix \ref{app:dp_if}), the influence function converges to zero faster when $G$ is lighter.
Thus, since the standard normal distribution is lighter than the standard logistic distribution, the first row of Figure \ref{fg:plot_if} converges to zero faster than the second row.

For the influence functions with the DP and $\gamma$-divergences, when the tuning parameters $\alpha$ and $\gamma$ are the same value, the influence function with the $\gamma$-divergence takes smaller values when $x$ is around $0$, but when $|x|\to \infty$, the influence functions with the DP and $\gamma$-divergences take the same value.

\section{Asymptotic properties for proposed robust estimators and their test procedures}
\label{sec:asymp}
This section presents the asymptotic distributions of the proposed DP and $\gamma$-estimators and proposes the test procedures using them.

\subsection{Case: Density-Power divergence}
\label{sec:dp_asym}
Under general regularity conditions (\citealp{huber2009robust}), the proposed DP-estimator $\hat{\bm{\theta}}_{DP}$ \eqref{eq:dp_est} is asymptotically a multivariate normal distribution, i.e.,
\begin{equation}
\label{eq:dp_asym}
\sqrt{n} (\hat{\bm{\theta}}_{DP} - \bm{\theta}) \overset{d}{\to} N(\bm{0}, \bm{V}_{DP}(\bm{\theta}, \psi_{DP})),
\end{equation}
where the asymptotic covariance matrix is
\begin{equation*}
\bm{V}_{DP}(\bm{\theta}, \psi_{DP}) = \left( \bm{M}_{DP}(\bm{\theta}, \psi_{DP}) \right)^{-1} \bm{Q}_{DP}(\bm{\theta}, \psi_{DP}) \left( \bm{M}_{DP}(\bm{\theta}, \psi_{DP}) \right)^{-1}
\end{equation*}
with
\begin{equation*}
\bm{M}_{DP}(\bm{\theta}, \psi_{DP}) = - \E_{(y,\bm{x})} \left[ \frac{\pd}{\pd \bm{\theta}} \psi_{DP}(y|\bm{x};\bm{\theta}) \right],
\end{equation*}
\begin{equation*}
\psi_{DP}(y|\bm{x};\bm{\theta}) = f(y|\bm{x};\bm{\theta})^{\alpha} s(y|\bm{x};\bm{\theta}) - \E_{(y|\bm{x})}[ f(y|\bm{x};\bm{\theta})^{\alpha} s(y|\bm{x};\bm{\theta}) ],
\end{equation*}
and
\begin{equation*}
\bm{Q}_{DP}(\bm{\theta}, \psi_{DP}) = \E_{(y,\bm{x})} \left[ \psi_{DP}(y|\bm{x};\bm{\theta}) \psi_{DP}(y|\bm{x};\bm{\theta})^\top \right].
\end{equation*}
Since the estimator $\hat{\bm{V}}_{DP}(\bm{\theta}, \psi_{DP})$ of the asymptotic covariance matrix $\bm{V}_{DP}(\bm{\theta}, \psi_{DP})$ is
\begin{equation*}
\hat{\bm{V}}_{DP}(\bm{\theta}, \psi_{DP}) = \left( \hat{\bm{M}}_{DP}(\bm{\theta}, \psi_{DP}) \right)^{-1} \hat{\bm{Q}}_{DP}(\bm{\theta}, \psi_{DP}) \left( \hat{\bm{M}}_{DP}(\bm{\theta}, \psi_{DP}) \right)^{-1},
\end{equation*}
where
\begin{equation*}
\hat{\bm{M}}_{DP}(\bm{\theta}, \psi_{DP}) = - \frac{1}{n} \sum_{i=1}^n \left. \frac{\pd}{\pd \bm{\theta}} \psi_{DP}(y_i|\bm{x}_i;\bm{\theta}) \right|_{\bm{\theta}=\hat{\bm{\theta}}_{DP}}
\end{equation*}
and
\begin{equation*}
\hat{\bm{Q}}_{DP}(\bm{\theta}, \psi_{DP}) = \frac{1}{n} \sum_{i=1}^n \psi_{DP}(y_i|\bm{x}_i;\hat{\bm{\theta}}_{DP}) \psi_{DP}(y_i|\bm{x}_i;\hat{\bm{\theta}}_{DP})^\top,
\end{equation*}
and $\hat{\bm{V}}_{DP}(\bm{\theta}, \psi_{DP})$ is consistent for $\bm{V}_{DP}(\bm{\theta}, \psi_{DP})$, then with the equation \eqref{eq:dp_asym}, $\sqrt{n} (\hat{\bm{\theta}}_{DP}-\bm{\theta})$ is asymptotically a multivariate normal distribution $N(\bm{0}, \hat{\bm{V}}_{DP}(\bm{\theta}, \psi_{DP}))$.

Hence under the null hypothesis $H_0^k: \beta_k = 0$ for $k=1,2,\ldots,p$, 
\begin{equation*}
T_{\beta_k} = \frac{ \hat{\beta}_k^{DP} - \beta_k }{ \sqrt{ \frac{ \hat{\sigma}_{DP}^2[\beta_k] }{n} } } \overset{d}{\to} N(0,1),
\end{equation*}
where $\hat{\sigma}_{DP}^2[\beta_k]$ is the $k$-th element on the diagonal of $\hat{\bm{V}}_{DP}(\bm{\beta}, \psi_{DP})$, $\hat{\bm{V}}_{DP}(\bm{\beta}, \psi_{DP})$ is the submatrix of $\hat{\bm{V}}_{DP}(\bm{\theta}, \psi_{DP})$ related to the coefficient parameters $\bm{\beta}$, and $\hat{\beta}_k^{DP}$ is the $k$-th element of the proposed DP-estimator $\hat{\bm{\beta}}_{DP}$ of $\bm{\beta}$.

\subsection{Case: $\gamma$-divergence}
Under general regularity conditions (\citealp{huber2009robust}), the proposed $\gamma$-estimator $\hat{\bm{\theta}}_{\gamma}$ \eqref{eq:dp_est} is asymptotically a multivariate normal distribution, i.e.,
\begin{equation}
\label{eq:g_asym}
\sqrt{n} (\hat{\bm{\theta}}_{\gamma} - \bm{\theta}) \overset{d}{\to} N(\bm{0}, \bm{V}_{\gamma}(\bm{\theta}, \psi_{\gamma})),
\end{equation}
where the asymptotic covariance matrix is
\begin{equation*}
\bm{V}_{\gamma}(\bm{\theta}, \psi_{\gamma}) = \left( \bm{M}_{\gamma}(\bm{\theta}, \psi_{\gamma}) \right)^{-1} \bm{Q}_{\gamma}(\bm{\theta}, \psi_{\gamma}) \left( \bm{M}_{\gamma}(\bm{\theta}, \psi_{\gamma}) \right)^{-1}
\end{equation*}
with
\begin{equation*}
\bm{M}_{\gamma}(\bm{\theta}, \psi_{\gamma}) = - \E_{(y,\bm{x})} \left[ \frac{\pd}{\pd \bm{\theta}} \psi_{\gamma}(y|\bm{x};\bm{\theta}) \right],
\end{equation*}
\begin{equation*}
\psi_{\gamma}(y|\bm{x};\bm{\theta}) = f(y|\bm{x};\bm{\theta})^{\gamma} s(y|\bm{x};\bm{\theta}) \E_{(y|\bm{x})}[ f(y|\bm{x};\bm{\theta})^{\gamma} ] - f(y|\bm{x};\bm{\theta})^{\gamma} \E_{(y|\bm{x})}[ f(y|\bm{x};\bm{\theta})^{\gamma} s(y|\bm{x};\bm{\theta}) ],
\end{equation*}
and
\begin{equation*}
\bm{Q}_{\gamma}(\bm{\theta}, \psi_{\gamma}) = \E_{(y,\bm{x})} \left[ \psi_{\gamma}(y|\bm{x};\bm{\theta}) \psi_{\gamma}(y|\bm{x};\bm{\theta})^\top \right].
\end{equation*}
Since the estimator $\hat{\bm{V}}_{\gamma}(\bm{\theta}, \psi_{\gamma})$ of the asymptotic covariance matrix $\bm{V}_{\gamma}(\bm{\theta}, \psi_{\gamma})$ is
\begin{equation*}
\hat{\bm{V}}_{\gamma}(\bm{\theta}, \psi_{\gamma}) = \left( \hat{\bm{M}}_{\gamma}(\bm{\theta}, \psi_{\gamma}) \right)^{-1} \hat{\bm{Q}}_{\gamma}(\bm{\theta}, \psi_{\gamma}) \left( \hat{\bm{M}}_{\gamma}(\bm{\theta}, \psi_{\gamma}) \right)^{-1},
\end{equation*}
where
\begin{equation*}
\hat{\bm{M}}_{\gamma}(\bm{\theta}, \psi_{\gamma}) = - \frac{1}{n} \sum_{i=1}^n \left. \frac{\pd}{\pd \bm{\theta}} \psi_{\gamma}(y_i|\bm{x}_i;\bm{\theta}) \right|_{\bm{\theta}=\hat{\bm{\theta}}_{\gamma}}
\end{equation*}
and
\begin{equation*}
\hat{\bm{Q}}_{\gamma}(\bm{\theta}, \psi_{\gamma}) = \frac{1}{n} \sum_{i=1}^n \psi_{\gamma}(y_i|\bm{x}_i;\hat{\bm{\theta}}_{\gamma}) \psi_{\gamma}(y_i|\bm{x}_i;\hat{\bm{\theta}}_{\gamma})^\top,
\end{equation*}
and $\hat{\bm{V}}_{\gamma}(\bm{\theta}, \psi_{\gamma})$ is consistent for $\bm{V}_{\gamma}(\bm{\theta}, \psi_{\gamma})$, then with the equation \eqref{eq:g_asym}, $\sqrt{n} (\hat{\bm{\theta}}_{\gamma}-\bm{\theta})$ is asymptotically a multivariate normal distribution $N(\bm{0}, \hat{\bm{V}}_{\gamma}(\bm{\theta}, \psi_{\gamma}))$.

Hence under the null hypothesis $H_0^k: \beta_k = 0$ for $k=1,2,\ldots,p$, 
\begin{equation*}
T_{\beta_k} = \frac{ \hat{\beta}_k^{\gamma} - \beta_k }{ \sqrt{ \frac{ \hat{\sigma}_{\gamma}^2[\beta_k] }{n} } } \overset{d}{\to} N(0,1),
\end{equation*}
where $\hat{\sigma}_{\gamma}^2[\beta_k]$ is the $k$-th element on the diagonal of $\hat{\bm{V}}_{\gamma}(\bm{\beta}, \psi_{\gamma})$, $\hat{\bm{V}}_{\gamma}(\bm{\beta}, \psi_{\gamma})$ is the submatrix of $\hat{\bm{V}}_{\gamma}(\bm{\theta}, \psi_{\gamma})$ related to the coefficient parameters $\bm{\beta}$, and $\hat{\beta}_k^{\gamma}$ is the $k$-th element of the proposed $\gamma$-estimator $\hat{\bm{\beta}}_{\gamma}$ of $\bm{\beta}$.

\section{Numerical Experiments}
\label{sec:ne}
This section shows the performance of the proposed robust ordinal response model with the DP and $\gamma$-divergences by means of numerical experiments with artificially generated data and two real data.

\subsection{Simulation study}
This section performs some simulation studies of the proposed methods, referring to Section 5 of \cite{scalera2021robust}.
We consider the following latent variable model \eqref{eq:lvm} for ordered categorical data $y$ with five categories (i.e., $M=5$) as the response variable.
\begin{equation*}
z = x\beta_1 + d\beta_2 + xd\beta_3 + \ep,
\end{equation*}
where $x \sim N(0,1)$, $d \sim {\rm Bernoulli}(0.25)$, $x$ and $d$ are mutually independent.
The random variable $\ep$ has standard normal, logistic, and Gumbel distributions, and in this numerical experiment we use the link function suitable for the each error distribution.
That is, probit link for normal distribution, logit link for logistic distribution, and log-log link for Gumbel distribution are used.
This is because the misspecification of the link function causes a substantial bias in the parameter estimation, apart from the effect of outliers on the inference.
The development of an inference method that is robust to both outliers and the misspecification of link functions is the future work.
The true values of the regression coefficients $\bm{\beta}$ and cutoff $\bm{\delta}$ parameters are set as follows according to the distribution $\ep$ has.
\begin{itemize}
\item When $\ep$ has the normal distribution:
\begin{equation*}
(\beta_1,\beta_2,\beta_3)=(2.5,1.2,0.7) ~~ \mbox{and} ~~ (\delta_1,\delta_2,\delta_3,\delta_4)=(-3.0,-0.7,1.6,3.9).
\end{equation*}
\item When $\ep$ has the logistic distribution:
\begin{equation*}
(\beta_1,\beta_2,\beta_3)=(2.5,1.2,0.7) ~~ \mbox{and} ~~ (\delta_1,\delta_2,\delta_3,\delta_4)=(-3.3,-0.8,1.7,4.2).
\end{equation*}
\item When $\ep$ has the Gumbel distribution:
\begin{equation*}
(\beta_1,\beta_2,\beta_3)=(2.5,1.2,0.7) ~~ \mbox{and} ~~ (\delta_1,\delta_2,\delta_3,\delta_4)=(-2.9,1.0,2.9,4.8).
\end{equation*}
\end{itemize}
We generate 200 observations with the above settings, and then generate outliers in the covariate $x_i$ for 5 and 10 percent of the total observations (i.e., 10 and 20 outliers) from $N(20,1)$ and replace them with $x_i$ at random.

Using the link functions corresponding to the error distributions of $\ep$, we compare the maximum likelihood method with the proposed methods using the DP and $\gamma$-divergences, using bias, mean squared error (MSE), and percentage of correctly classified responses (CCR): 
\begin{align*}
{\rm Bias}_j = \frac{1}{S} \sum_{s=1}^S \hat{\theta}_{js} - \theta_{j}, ~~ {\rm MSE}_j = \frac{1}{S} \sum_{s=1}^S (\hat{\theta}_{js} - \theta_{j})^2, ~~ {\rm CCR} = \frac{1}{S} \sum_{s=1}^S I(\hat{y}_{s} = y_{s}^*),
\end{align*}
where $S$ is the number of iterations ($S=1000$ in this simulation), $j$ is the index for each parameter, $\hat{\theta}_{js}$ is the estimated value of the parameter at the $s$th iteration, $\hat{y}_s$ is the predicted ordinal response using $\hat{\theta}_{js}$ and the newly generated non-outlier covariates $x$ and $d$ at the $s$th iteration, $y_s^*$ is the validation data at the $s$th iteration, and $I(A)$ is an indicator function that returns 1 when $A$ is true and 0 otherwise.

The maximum likelihood method using the cauchit link, which is the distribution function of the Cauchy distribution and the link function whose influence function is bounded in  shown by \cite{scalera2021robust}, is also included in the comparison.
This is because as mentioned earlier, the misspecification of the link function certainly causes a substantial bias in the parameter estimation, but as shown in our simulation results to be presented later, it seems to perform better in the prediction than using a link function whose influence function is not bounded.

Tables \ref{tb:ne-probit} to \ref{tb:ne-loglog} summarize the simulation results for the settings described above.
Tables \ref{tb:ne-probit}, \ref{tb:ne-logit}, and \ref{tb:ne-loglog} show the results when $\ep$ has the normal distribution, logistic distribution, and Gumbel distribution, respectively.
When the outlier ratio is 0$\%$, the maximum likelihood methods with suitable links minimize bias and MSE and have high CCR values in most cases, no matter which distribution $\ep$ follows.
Notably, however, the proposed methods with the DP and $\gamma$-divergences also perform as well as the maximum likelihood methods.
Since it is difficult to determine whether there are outliers or not in most cases in real data, it is a very desirable property that the estimation accuracy is equivalent to that of the maximum likelihood method even when there are no outliers in reality.
When the cauchit link is used in the maximum likelihood method, the bias and MSE are still large due to the misspecification of the link function, but CCR is not that much worse than other methods.

When the data contain outliers, this simulation suggests that the maximum likelihood methods are considerably affected by the outliers.
In addition, the maximum likelihood method with the cauchit link misspecifies the link function, resulting in considerably larger values of bias and MSE.
On the other hand, as \cite{scalera2021robust} also mentioned, since the influence function in the maximum likelihood method with the cauchit link is bounded, the prediction accuracy in terms of CCR values is better than that of the maximum likelihood method with correctly specified links.
However, it can be seen from Tables \ref{tb:ne-probit} to \ref{tb:ne-loglog} that our proposed methods are more accurate in terms of bias, MSE, and CCR.
It is particularly noteworthy that the maximum likelihood method with the cauchit link gives considerably worse CCR values when the outlier ratio increases, while our proposed methods show almost no decrease in the CCR values.
This may be because, as shown in Section \ref{sec:if}, the proposed methods whose influence functions satisfy not only boundedness but also redescendence are more robust than the maximum likelihood method with the cauchit link.

Comparing the methods using the DP and $\gamma$-divergences, it is found that there is little difference in all the values of the bias, MSE, and CCR.
However, although the difference is really small, the values of the bias and MSE are smaller for the method using the DP divergence, while the value of the CCR is larger for the method using $\gamma$-divergences.
This tendency may slightly change depending on the selection of the tuning parameters, $\alpha$ and $\gamma$, as described in Remark below, but our empirical results are generally good when the values of $\alpha$ and $\gamma$ are set at around $0.3\sim0.5$.

%

Next, we conduct a numerical experiment that is almost the same as the data generation process described above, but this time the outlier ratio in the covariate $x_i$ is fixed at 5\% and the mean of the normal distribution of outliers is varied.
Tables \ref{tb:ne2-probit}, \ref{tb:ne2-logit}, and \ref{tb:ne2-loglog} show the results for the mean parameter values of 5, 10, and 20 with the probit, logit, and log-log link functions.
When the value of the mean parameter is set to 20, the settings are the same as those in the middle panels of Tables \ref{tb:ne-probit} to \ref{tb:ne-loglog}, and hence these values are reproduced in the tables.
The results in Tables \ref{tb:ne2-probit} to \ref{tb:ne2-loglog} are consistent with the results of the theorems on the influence function in Section \ref{sec:if}.
Since the maximum likelihood method with the probit, logit, and log-log links does not satisfy boundedness, the accuracy of inference on the bias, MSE, and CCR becomes worse for larger values of outliers.
In contrast, the influence function of our method with the DP and $\gamma$-divergences satisfies both boundedness and redescendence, and the inference accuracy is robust to increasing values of outliers.
Moreover, since our method can ignore the influence of outliers more as the value of outliers increases, the inference accuracy is better for most of the indices when the value of the mean parameter of the contaminated normal distribution is 20 than for the other cases.

\begin{rem}[Selection of tuning parameters, $\alpha$ and $\gamma$]
The DP and $\gamma$-divergences use tuning parameters $\alpha$ and $\gamma$, respectively, to control the robustness against outliers.
\cite{basu1998robust} mentioned that if the value of tuning parameter is smaller than necessary, the effect of outliers on the estimators obtained by using these divergences cannot be adequately removed, on the contrary, if a larger tuning parameter value is used than necessary, the statistical efficiency of the estimators will be lost instead.
Namely, there is a trade-off between robustness and efficiency.
Therefore, we would like to lightly discuss how to choose tuning parameters.

A simple and well-known method for selecting tuning parameters is to use the asymptotic relative efficiency, which is calculated using the asymptotic variance of the estimator in the robust divergence-based method and the asymptotic variance of the estimator in the maximum likelihood method.
Another well-known method is the method using cross-validation, but it is not straightforward when outliers exist in the data, so we have to find appropriate values of tuning parameters through trial and error.
Recently, \cite{sugasawa2021selection} improved the method proposed by \cite{warwick2005choosing} and \cite{basak2021optimal} for selecting tuning parameters based on the asymptotic approximation of the mean square error in a simple normal model and linear regression, and proposed a method for selecting tuning parameters based on the asymptotic approximation of the Hyv{\"a}rinen score (\citealp{shao2019bayesian, dawid2015bayesian}) using unnormalized models based on robust divergence.
The extension of their method to the ordinal response model will be the subject of future work.
\end{rem}

\begin{landscape}
\begin{table}[t]
\caption{Bias, mean square error (MSE), and percentage of correctly classified responses (CCR) using the maximum likelihood, DP and $\gamma$-divergences methods in the ordinal response model with the probit link, and maximum likelihood method with the cauchit link.
The outlier ratios are 0, 5, and 10\%.
Bold, *, and ** denote the first, second, and third best performance, respectively.}
\begin{center}
\scalebox{0.78}{
\begin{tabular}{rc lllllll c lllllll c l} \hline
& & \multicolumn{7}{l}{Bias} & & \multicolumn{7}{l}{MSE} & & CCR \\ \cline{3-9} \cline{11-17} \cline{19-19}
& tuning & $\beta_1$ & $\beta_2$ & $\beta_3$ & $\delta_1$ & $\delta_2$ & $\delta_3$ & $\delta_4$ & & $\beta_1$ & $\beta_2$ & $\beta_3$ & $\delta_1$ & $\delta_2$ & $\delta_3$ & $\delta_4$ & & \\ \hline 
Outlier Ratio 0\%&$$&$$&$$&$$&$$&$$&$$&$$&$$&$$&$$&$$&$$&$$&$$&$$&$$&$$\tabularnewline
ML&$$&${\bf 0.0767}$&${\bf 0.0350}$&${\bf 0.0435}$&${\bf 0.0916}$&${\bf 0.0216}$&${\bf 0.0457}$&${\bf 0.1219}$&$$&$ {\bf 0.0596}$&${\bf 0.0597}$&$ {\bf 0.0993}$&$ {\bf 0.0996}$&${\bf 0.0276}$&${\bf 0.0450}$&$ {\bf 0.1552}$&$$&$0.6854$\tabularnewline
ML $+$ Cauchit&$$&$3.8257$&$1.8362$&$1.1407$&$4.8149$&$1.0724$&$2.4343$&$6.2213$&$$&$15.6089$&$3.9851$&$ 2.1628$&$24.7377$&$1.4026$&$6.4745$&$41.2110$&$$&$0.6827$\tabularnewline
DP&$0.3$&$0.0825$*&$0.0373$*&$0.0439$*&$0.0990$*&$0.0237$*&$0.0495$*&$0.1301$*&$$&$ 0.0682$*&$0.0658$*&$ 0.1072$*&$ 0.1137$*&$0.0301$*&$0.0496$*&$ 0.1774$*&$$&$0.6859$**\tabularnewline
DP&$0.5$&$0.1018$&$0.0468$&$0.0518$&$0.1218$&$0.0288$&$0.0612$&$0.1603$&$$&$ 0.0848$&$0.0759$&$ 0.1210$&$ 0.1388$&$0.0336$&$0.0579$&$ 0.2187$&$$&${\bf 0.6862}$\tabularnewline
$\gamma$&$0.3$&$0.0829$**&$0.0375$**&$0.0440$**&$0.0995$**&$0.0238$**&$0.0498$**&$0.1309$**&$$&$ 0.0687$**&$0.0659$**&$ 0.1073$**&$ 0.1144$**&$0.0302$**&$0.0498$**&$ 0.1788$**&$$&$0.6859$**\tabularnewline
$\gamma$&$0.5$&$0.1045$&$0.0481$&$0.0526$&$0.1248$&$0.0296$&$0.0628$&$0.1643$&$$&$ 0.0877$&$0.0767$&$ 0.1217$&$ 0.1428$&$0.0340$&$0.0591$&$ 0.2256$&$$&${\bf 0.6862}$\tabularnewline
\multicolumn{19}{l}{} \\
Outlier Ratio 5\%&$$&$$&$$&$$&$$&$$&$$&$$&$$&$$&$$&$$&$$&$$&$$&$$&$$&$$\tabularnewline
ML&$$&$2.4694$&$0.6027$&$0.9202$&$1.7694$&$0.4388$&$0.8950$&$2.1682$&$$&$ 6.0983$&$0.4111$&$ 0.9292$&$ 3.1509$&$0.2043$&$0.8152$&$ 4.7388$&$$&$0.4162$\tabularnewline
ML $+$ Cauchit&$$&$0.1015$&$0.9756$&$2.2620$&$1.5232$&$0.1972$&$0.5918$&$3.2111$&$$&$ 3.7708$&$6.7094$&$24.7913$&$ 4.2271$&$0.2695$&$1.2385$&$44.8350$&$$&$0.5779$\tabularnewline
DP&$0.3$&$0.0180$*&${\bf 0.0008}$&${\bf 0.0277}$&$0.0039$*&${\bf 0.0003}$&$0.0069$*&${\bf 0.0036}$&$$&$ {\bf 0.0538}$&${\bf 0.0645}$&$ {\bf 0.1004}$&$ {\bf 0.0948}$&${\bf 0.0284}$&${\bf 0.0417}$&$ {\bf 0.1395}$&$$&$0.6829$\tabularnewline
DP&$0.5$&${\bf 0.0086}$&$0.0056$*&$0.0306$*&${\bf 0.0035}$&$0.0039$*&${\bf 0.0041}$&$0.0175$*&$$&$ 0.0637$**&$0.0722$**&$ 0.1114$**&$ 0.1094$**&$0.0313$**&$0.0469$**&$ 0.1640$**&$$&${\bf 0.6834}$\tabularnewline
$\gamma$&$0.3$&$0.0207$**&$0.0196$**&$0.0392$**&$0.0410$**&$0.0110$**&$0.0178$**&$0.0625$**&$$&$ 0.0566$*&$0.0673$*&$ 0.1047$*&$ 0.1014$*&$0.0297$*&$0.0440$*&$ 0.1508$*&$$&$0.6832$*\tabularnewline
$\gamma$&$0.5$&$0.0802$&$0.0488$&$0.0574$&$0.1059$&$0.0286$&$0.0523$&$0.1524$&$$&$ 0.0794$&$0.0816$&$ 0.1239$&$ 0.1364$&$0.0356$&$0.0558$&$ 0.2112$&$$&$0.6830$**\tabularnewline
\multicolumn{19}{l}{} \\
Outlier Ratio 10\%&$$&$$&$$&$$&$$&$$&$$&$$&$$&$$&$$&$$&$$&$$&$$&$$&$$&$$\tabularnewline
ML&$$&$2.4863$&$0.6091$&$0.9294$&$1.7779$&$0.4362$&$0.9058$&$2.1811$&$$&$ 6.1817$&$0.4116$&$ 0.9308$&$ 3.1810$&$0.2029$&$0.8354$&$ 4.7976$&$$&$0.4211$\tabularnewline
ML $+$ Cauchit&$$&$2.3053$&$0.0783$&$2.5291$&${\bf 0.0143}$&$0.2849$&$0.4374$&$0.6278$&$$&$ 5.8831$&$0.7122$&$ 8.1479$&$ 0.5887$&$0.1366$&$0.3513$&$ 3.7515$&$$&$0.4341$\tabularnewline
DP&$0.3$&${\bf 0.0574}$&${\bf 0.0326}$&${\bf 0.0346}$&$0.0734$*&${\bf 0.0179}$&${\bf 0.0359}$&${\bf 0.0960}$&$$&$ {\bf 0.0701}$&${\bf 0.0710}$&$ {\bf 0.1136}$&$ {\bf 0.1163}$&${\bf 0.0323}$&${\bf 0.0508}$&$ {\bf 0.1778}$&$$&${\bf 0.6805}$\tabularnewline
DP&$0.5$&$0.0744$*&$0.0420$*&$0.0423$*&$0.0919$**&$0.0229$*&$0.0459$*&$0.1228$*&$$&$ 0.0858$*&$0.0815$**&$ 0.1277$**&$ 0.1384$*&$0.0358$*&$0.0587$*&$ 0.2163$*&$$&$0.6794$\tabularnewline
$\gamma$&$0.3$&$0.1375$**&$0.0714$**&$0.0580$**&$0.1662$&$0.0401$**&$0.0869$**&$0.2177$**&$$&$ 0.0920$**&$0.0803$*&$ 0.1237$*&$ 0.1495$**&$0.0364$**&$0.0617$**&$ 0.2329$**&$$&$0.6803$*\tabularnewline
$\gamma$&$0.5$&$0.2601$&$0.1322$&$0.0978$&$0.3069$&$0.0744$&$0.1641$&$0.4053$&$$&$ 0.1704$&$0.1134$&$ 0.1596$&$ 0.2612$&$0.0489$&$0.0984$&$ 0.4244$&$$&$0.6795$**\tabularnewline
\hline
\end{tabular}
}
\end{center}
\label{tb:ne-probit}
\end{table}
\end{landscape} %

\begin{landscape}
\begin{table}[t]
\caption{Bias, mean square error (MSE), and percentage of correctly classified responses (CCR) using the maximum likelihood, DP and $\gamma$-divergences methods in the ordinal response model with the logit link, and maximum likelihood method with the cauchit link.
The outlier ratios are 0, 5, and 10\%.
Bold, *, and ** denote the first, second, and third best performance, respectively.}
\begin{center}
\scalebox{0.78}{
\begin{tabular}{rc lllllll c lllllll c l} \hline
& & \multicolumn{7}{l}{Bias} & & \multicolumn{7}{l}{MSE} & & CCR \\ \cline{3-9} \cline{11-17} \cline{19-19}
& tuning & $\beta_1$ & $\beta_2$ & $\beta_3$ & $\delta_1$ & $\delta_2$ & $\delta_3$ & $\delta_4$ & & $\beta_1$ & $\beta_2$ & $\beta_3$ & $\delta_1$ & $\delta_2$ & $\delta_3$ & $\delta_4$ & & \\ \hline 
Outlier Ratio 0\%&$$&$$&$$&$$&$$&$$&$$&$$&$$&$$&$$&$$&$$&$$&$$&$$&$$&$$\tabularnewline
ML&$$&${\bf 0.0578}$&${\bf 0.0310}$&${\bf 0.0390}$&${\bf 0.0749}$&${\bf 0.0125}$&${\bf 0.0372}$&${\bf 0.0976}$&$$&${\bf 0.0695}$&${\bf 0.1263}$&${\bf 0.1836}$&${\bf 0.1311}$&${\bf 0.0501}$&${\bf 0.0654}$&${\bf 0.1791}$&$$&$0.5441$\tabularnewline
ML $+$ Cauchit&$$&$0.4540$&$0.2302$&$0.1798$&$0.8128$&$0.1393$&$0.2939$&$1.0295$&$$&$0.3846$&$0.2805$&$0.3500$&$1.0078$&$0.1010$&$0.2158$&$1.5714$&$$&$0.5397$\tabularnewline
DP&$0.3$&$0.0713$*&$0.0364$*&$0.0414$*&$0.0906$*&$0.0164$*&$0.0445$*&$0.1191$*&$$&$0.0794$*&$0.1366$*&$0.1963$*&$0.1462$*&$0.0526$*&$0.0705$*&$0.2043$*&$$&$0.5450$\tabularnewline
DP&$0.5$&$0.0887$&$0.0447$&$0.0486$&$0.1115$&$0.0212$&$0.0546$&$0.1474$&$$&$0.0935$&$0.1508$&$0.2151$&$0.1689$&$0.0565$&$0.0781$&$0.2414$&$$&${\bf 0.5455}$\tabularnewline
$\gamma$&$0.3$&$0.0718$**&$0.0367$**&$0.0416$**&$0.0912$**&$0.0165$**&$0.0448$**&$0.1199$**&$$&$0.0800$**&$0.1368$**&$0.1965$**&$0.1470$**&$0.0527$**&$0.0707$**&$0.2057$**&$$&$0.5451$**\tabularnewline
$\gamma$&$0.5$&$0.0922$&$0.0465$&$0.0498$&$0.1154$&$0.0221$&$0.0568$&$0.1525$&$$&$0.0970$&$0.1523$&$0.2167$&$0.1736$&$0.0569$&$0.0796$&$0.2494$&$$&${\bf 0.5455}$\tabularnewline
\multicolumn{19}{l}{} \\
Outlier Ratio 5\%&$$&$$&$$&$$&$$&$$&$$&$$&$$&$$&$$&$$&$$&$$&$$&$$&$$&$$\tabularnewline
ML&$$&$2.4412$&$0.3784$&$1.4846$&$1.3235$&$0.3769$&$0.6316$&$1.4210$&$$&$5.9605$&$0.2196$&$2.2996$&$1.8165$&$0.1727$&$0.4368$&$2.1277$&$$&$0.3819$\tabularnewline
ML $+$ Cauchit&$$&$1.1169$&$0.0661$&$0.9473$&$0.2000$&$0.1815$&$0.3024$&$0.0674$&$$&$2.5090$&$0.8044$&$2.9639$&$0.5394$&$0.1182$&$0.2825$&$1.9291$&$$&$0.4696$\tabularnewline
DP&$0.3$&${\bf 0.0614}$&${\bf 0.0272}$&${\bf 0.0454}$&${\bf 0.0832}$&${\bf 0.0130}$&${\bf 0.0404}$&$0.1050$*&$$&${\bf 0.0778}$&${\bf 0.1424}$&${\bf 0.2146}$&${\bf 0.1431}$&${\bf 0.0543}$&${\bf 0.0723}$&${\bf 0.2073}$&$$&${\bf 0.5417}$\tabularnewline
DP&$0.5$&$0.0795$*&$0.0358$*&$0.0542$*&$0.1039$*&$0.0178$*&$0.0510$*&$0.1338$**&$$&$0.0916$**&$0.1568$**&$0.2357$**&$0.1643$**&$0.0586$**&$0.0801$**&$0.2437$**&$$&$0.5401$\tabularnewline
$\gamma$&$0.3$&$0.1081$**&$0.0498$**&$0.0598$**&$0.1343$**&$0.0258$**&$0.0701$**&$0.1733$&$$&$0.0893$*&$0.1500$*&$0.2252$*&$0.1616$*&$0.0576$*&$0.0793$*&$0.2371$*&$$&$0.5416$*\tabularnewline
$\gamma$&$0.5$&$0.1897$&$0.0891$&$0.0892$&$0.2244$&$0.0479$&$0.1207$&$0.2950$&$$&$0.1336$&$0.1803$&$0.2671$&$0.2265$&$0.0680$&$0.1028$&$0.3477$&$$&$0.5412$**\tabularnewline
\multicolumn{19}{l}{} \\
Outlier Ratio 10\%&$$&$$&$$&$$&$$&$$&$$&$$&$$&$$&$$&$$&$$&$$&$$&$$&$$&$$\tabularnewline
ML&$$&$2.4736$&$0.3798$&$1.5006$&$1.3333$&$0.3816$&$0.6414$&$1.4402$&$$&$6.1190$&$0.2191$&$2.3445$&$1.8466$&$0.1781$&$0.4505$&$2.1787$&$$&$0.3745$\tabularnewline
ML $+$ Cauchit&$$&$2.4430$&$0.2876$&$1.7258$&$0.7887$&$0.4208$&$0.7367$&$0.6024$&$$&$6.0382$&$0.4259$&$4.2623$&$0.9332$&$0.2084$&$0.6049$&$1.1041$&$$&$0.3735$\tabularnewline
DP&$0.3$&${\bf 0.0643}$&${\bf 0.0325}$&${\bf 0.0438}$&${\bf 0.0887}$&${\bf 0.0157}$&${\bf 0.0387}$&${\bf 0.1107}$&$$&${\bf 0.0862}$&${\bf 0.1515}$&${\bf 0.2242}$&${\bf 0.1554}$&${\bf 0.0582}$&${\bf 0.0762}$&${\bf 0.2184}$&$$&$0.5370$**\tabularnewline
DP&$0.5$&$0.0842$*&$0.0434$*&$0.0530$*&$0.1117$*&$0.0210$*&$0.0500$*&$0.1420$*&$$&$0.1022$*&$0.1682$*&$0.2462$*&$0.1797$*&$0.0628$*&$0.0846$*&$0.2587$*&$$&${\bf 0.5376}$\tabularnewline
$\gamma$&$0.3$&$0.1569$**&$0.0774$**&$0.0721$**&$0.1903$**&$0.0413$**&$0.0974$**&$0.2463$**&$$&$0.1143$**&$0.1688$**&$0.2462$**&$0.1992$**&$0.0655$**&$0.0919$**&$0.2887$**&$$&$0.5368$\tabularnewline
$\gamma$&$0.5$&$0.2998$&$0.1483$&$0.1207$&$0.3493$&$0.0807$&$0.1864$&$0.4589$&$$&$0.2102$&$0.2242$&$0.3122$&$0.3369$&$0.0846$&$0.1391$&$0.5194$&$$&$0.5375$*\tabularnewline
\hline
\end{tabular}
}
\end{center}
\label{tb:ne-logit}
\end{table}
\end{landscape} %

\begin{landscape}
\begin{table}[t]
\caption{Bias, mean square error (MSE), and percentage of correctly classified responses (CCR) using the maximum likelihood, DP and $\gamma$-divergences methods in the ordinal response model with the log-log link, and maximum likelihood method with the cauchit link.
The outlier ratios are 0, 5, and 10\%.
Bold, *, and ** denote the first, second, and third best performance, respectively.}
\begin{center}
\scalebox{0.78}{
\begin{tabular}{rc lllllll c lllllll c l} \hline
& & \multicolumn{7}{l}{Bias} & & \multicolumn{7}{l}{MSE} & & CCR \\ \cline{3-9} \cline{11-17} \cline{19-19}
& tuning & $\beta_1$ & $\beta_2$ & $\beta_3$ & $\delta_1$ & $\delta_2$ & $\delta_3$ & $\delta_4$ & & $\beta_1$ & $\beta_2$ & $\beta_3$ & $\delta_1$ & $\delta_2$ & $\delta_3$ & $\delta_4$ & & \\ \hline 
Outlier Ratio 0\%&$$&$$&$$&$$&$$&$$&$$&$$&$$&$$&$$&$$&$$&$$&$$&$$&$$&$$\tabularnewline
ML&$$&${\bf 0.0918}$&$0.0563$**&$0.0289$**&${\bf 0.1176}$&${\bf 0.0285}$&${\bf 0.0827}$&${\bf 0.1431}$&$$&${\bf 0.0763}$&${\bf 0.0814}$&$ 0.1196$**&$ {\bf 0.1424}$&${\bf 0.0394}$&${\bf 0.1151}$&$ {\bf 0.2664}$&$$&${\bf 0.6638}$\tabularnewline
ML $+$ Cauchit&$$&$2.5500$&$1.2345$&$0.7202$&$4.0935$&$0.1432$&$2.1380$&$4.2819$&$$&$7.1430$&$2.0246$&$ 1.1506$&$17.8962$&$0.2194$&$5.3678$&$20.3844$&$$&$0.6614$\tabularnewline
DP&$0.3$&$0.0963$*&${\bf 0.0517}$&${\bf 0.0266}$&$0.1234$*&$0.0321$*&$0.0937$*&$0.1628$*&$$&$0.0861$*&$0.0864$*&$ {\bf 0.1132}$&$ 0.1445$*&$0.0421$*&$0.1218$*&$ 0.2800$*&$$&$0.6632$**\tabularnewline
DP&$0.5$&$0.1192$&$0.0624$&$0.0348$&$0.1526$&$0.0402$&$0.1180$&$0.2042$&$$&$0.1082$&$0.0996$&$ 0.1285$&$ 0.1811$&$0.0475$&$0.1472$&$ 0.3454$&$$&$0.6637$*\tabularnewline
$\gamma$&$0.3$&$0.0974$**&$0.0522$*&$0.0268$*&$0.1247$**&$0.0326$**&$0.0950$**&$0.1647$**&$$&$0.0872$**&$0.0868$**&$ 0.1134$*&$ 0.1460$**&$0.0424$**&$0.1232$**&$ 0.2833$**&$$&$0.6632$**\tabularnewline
$\gamma$&$0.5$&$0.1248$&$0.0651$&$0.0361$&$0.1595$&$0.0427$&$0.1241$&$0.2136$&$$&$0.1141$&$0.1016$&$ 0.1298$&$ 0.1903$&$0.0489$&$0.1544$&$ 0.3624$&$$&$0.6632$**\tabularnewline
\multicolumn{19}{l}{} \\
Outlier Ratio 5\%&$$&$$&$$&$$&$$&$$&$$&$$&$$&$$&$$&$$&$$&$$&$$&$$&$$&$$\tabularnewline
ML&$$&$2.4717$&$0.2387$&$1.0820$&$2.0293$&$0.3192$&$1.1554$&$1.7542$&$$&$6.1098$&$0.0903$&$ 1.2414$&$ 4.1320$&$0.1209$&$1.3758$&$ 3.1793$&$$&$0.5044$\tabularnewline
ML $+$ Cauchit&$$&$0.8895$&$0.1734$&$2.6148$&$1.8459$&$0.5027$&$0.2019$&$1.5774$&$$&$4.1309$&$2.3968$&$29.9653$&$ 5.4273$&$0.4916$&$2.3754$&$28.1556$&$$&$0.5742$\tabularnewline
DP&$0.3$&${\bf 0.0649}$&${\bf 0.0439}$&${\bf 0.0296}$&${\bf 0.0951}$&${\bf 0.0271}$&${\bf 0.0665}$&${\bf 0.1138}$&$$&${\bf 0.0802}$&${\bf 0.0885}$&$ {\bf 0.1181}$&$ {\bf 0.1427}$&${\bf 0.0430}$&${\bf 0.1159}$&$ {\bf 0.2642}$&$$&$0.6684$*\tabularnewline
DP&$0.5$&$0.0877$*&$0.0555$*&$0.0377$*&$0.1225$*&$0.0358$*&$0.0906$*&$0.1542$*&$$&$0.1002$**&$0.1019$**&$ 0.1344$**&$ 0.1768$**&$0.0485$**&$0.1402$**&$ 0.3257$**&$$&$0.6682$**\tabularnewline
$\gamma$&$0.3$&$0.1159$*&$0.0683$*&$0.0431$*&$0.1573$*&$0.0516$*&$0.1245$*&$0.2012$*&$$&$0.0940$*&$0.0953$*&$ 0.1242$*&$ 0.1666$*&$0.0472$*&$0.1339$*&$ 0.3075$*&$$&$0.6681$\tabularnewline
$\gamma$&$0.5$&$0.2105$&$0.1145$&$0.0704$&$0.2741$&$0.0936$&$0.2286$&$0.3634$&$$&$0.1542$&$0.1249$&$ 0.1542$&$ 0.2682$&$0.0630$&$0.2087$&$ 0.4903$&$$&${\bf 0.6693}$\tabularnewline
\multicolumn{19}{l}{} \\
Outlier Ratio 10\%&$$&$$&$$&$$&$$&$$&$$&$$&$$&$$&$$&$$&$$&$$&$$&$$&$$&$$\tabularnewline
ML&$$&$2.4849$&$0.2352$&$1.0929$&$2.0422$&$0.3150$&$1.1556$&$1.7498$&$$&$6.1751$&${\bf 0.0881}$&$ 1.2570$&$ 4.1841$&$0.1180$&$1.3763$&$ 3.1656$&$$&$0.5092$\tabularnewline
ML $+$ Cauchit&$$&$2.3897$&$0.4202$&$2.6221$&$1.0840$&$0.7450$&$1.1649$&$0.2194$**&$$&$5.9613$&$0.4398$&$ 8.1162$&$ 2.0854$&$0.5887$&$1.5602$&$ 1.0103$&$$&$0.5086$\tabularnewline
DP&$0.3$&${\bf 0.0956}$&${\bf 0.0577}$&${\bf 0.0407}$&${\bf 0.1300}$&${\bf 0.0352}$&${\bf 0.0977}$&${\bf 0.1693}$&$$&${\bf 0.0933}$&$0.0966$*&$ {\bf 0.1293}$&$ {\bf 0.1599}$&${\bf 0.0464}$&${\bf 0.1332}$&$ {\bf 0.3132}$&$$&${\bf 0.6653}$\tabularnewline
DP&$0.5$&$0.1225$*&$0.0714$*&$0.0507$*&$0.1645$*&$0.0450$*&$0.1263$*&$0.2176$*&$$&$0.1187$*&$0.1121$**&$ 0.1500$**&$ 0.2045$*&$0.0527$*&$0.1624$*&$ 0.3945$*&$$&$0.6637$**\tabularnewline
$\gamma$&$0.3$&$0.1966$**&$0.1060$**&$0.0678$**&$0.2530$**&$0.0829$**&$0.2124$**&$0.3426$&$$&$0.1318$**&$0.1129$&$ 0.1435$*&$ 0.2232$**&$0.0564$**&$0.1825$**&$ 0.4346$**&$$&$0.6651$*\tabularnewline
$\gamma$&$0.5$&$0.3639$&$0.1872$&$0.1168$&$0.4636$&$0.1560$&$0.3975$&$0.6314$&$$&$0.2716$&$0.1692$&$ 0.1983$&$ 0.4569$&$0.0888$&$0.3531$&$ 0.8669$&$$&$0.6633$\tabularnewline
\hline
\end{tabular}
}
\end{center}
\label{tb:ne-loglog}
\end{table}
\end{landscape} %

\begin{landscape}
\begin{table}[t]
\caption{Bias, mean square error (MSE), and percentage of correctly classified responses (CCR) using the maximum likelihood, DP and $\gamma$-divergences methods in the ordinal response model with the probit link, and maximum likelihood method with the cauchit link.
The values of the means of the contaminated normal distribution are 5, 10, and 20.
Bold, *, and ** denote the first, second, and third best performance, respectively.}
\begin{center}
\scalebox{0.78}{
\begin{tabular}{rc lllllll c lllllll c l} \hline
& & \multicolumn{7}{l}{Bias} & & \multicolumn{7}{l}{MSE} & & CCR \\ \cline{3-9} \cline{11-17} \cline{19-19}
& tuning & $\beta_1$ & $\beta_2$ & $\beta_3$ & $\delta_1$ & $\delta_2$ & $\delta_3$ & $\delta_4$ & & $\beta_1$ & $\beta_2$ & $\beta_3$ & $\delta_1$ & $\delta_2$ & $\delta_3$ & $\delta_4$ & & \\ \hline 
Mean $=$ 5&$$&$$&$$&$$&$$&$$&$$&$$&$$&$$&$$&$$&$$&$$&$$&$$&$$&$$\tabularnewline
ML&$$&$2.0440$&$0.5313$&$0.7193$&$1.5244$&${\bf 1.2441}$&$2.0080$&$2.7121$&$$&$4.1870$&$0.3184$&$0.5875$&$ 2.3477$&${\bf 1.5589}$&$4.0463$&$ 7.3972$&$$&$0.5117$\tabularnewline
ML $+$ Cauchit&$$&$2.2633$&$1.2328$&$1.0651$&$3.2475$&$2.3629$&${\bf 0.2718}$&$3.3824$&$$&$5.6709$&$2.1436$&$2.2141$&$11.4171$&$5.7229$&${\bf 0.3973}$&$14.0638$&$$&$0.6722$\tabularnewline
DP&$0.3$&${\bf 0.0622}$&${\bf 0.0322}$&${\bf 0.0424}$&${\bf 0.1787}$&$1.7158$*&$1.2638$&$0.7889$&$$&${\bf 0.0694}$&${\bf 0.0690}$&${\bf 0.1106}$&$ {\bf 0.1412}$&$2.9745$*&$1.6458$&$ 0.7958$&$$&$0.6778$*\tabularnewline
DP&$0.5$&$0.0871$*&$0.0434$*&$0.0485$*&$0.2057$*&$1.7225$**&$1.2503$&$0.7546$**&$$&$0.0855$*&$0.0792$**&$0.1241$**&$ 0.1719$**&$3.0010$**&$1.6193$&$ 0.7762$**&$$&$0.6770$\tabularnewline
$\gamma$&$0.3$&$0.1030$**&$0.0518$**&$0.0541$**&$0.2258$**&$1.7270$&$1.2381$**&$0.7272$*&$$&$0.0794$**&$0.0732$*&$0.1156$*&$ 0.1660$*&$3.0143$&$1.5838$**&$ 0.7115$*&$$&${\bf 0.6780}$\tabularnewline
$\gamma$&$0.5$&$0.1821$&$0.0892$&$0.0766$&$0.3154$&$1.7485$&$1.1904$*&${\bf 0.6104}$&$$&$0.1229$&$0.0931$&$0.1389$&$ 0.2487$&$3.0953$&$1.4806$*&$ {\bf 0.6113}$&$$&$0.6771$**\tabularnewline
&$$&$$&$$&$$&$$&$$&$$&$$&$$&$$&$$&$$&$$&$$&$$&$$&$$&$$\tabularnewline
Mean $=$ 10&$$&$$&$$&$$&$$&$$&$$&$$&$$&$$&$$&$$&$$&$$&$$&$$&$$&$$\tabularnewline
ML&$$&$2.3746$&$0.5805$&$0.8828$&$1.6631$&${\bf 1.2370}$&$2.1421$&$2.9733$&$$&$5.6398$&$0.4230$&${\bf 0.9311}$&$ 2.7867$&${\bf 1.5419}$&$4.6034$&$ 8.8831$&$$&$0.4304$\tabularnewline
ML $+$ Cauchit&$$&$2.0269$&$1.2731$&$1.2982$&$3.0887$&$2.3042$&${\bf 0.1542}$&$3.4987$&$$&$5.3225$&$3.9183$&$9.0887$&$10.9447$&$5.4748$&${\bf 0.5146}$&$25.8916$&$$&$0.6672$\tabularnewline
DP&$0.3$&${\bf 0.0672}$&${\bf 0.0373}$&${\bf 0.0276}$&${\bf 0.1818}$&$1.7191$*&$1.2607$&$0.7850$&$$&${\bf 0.0677}$&${\bf 0.0703}$&$0.1053$*&$ {\bf 0.1408}$&$2.9857$*&$1.6364$&$ 0.7797$&$$&$0.6828$*\tabularnewline
DP&$0.5$&$0.0863$*&$0.0480$*&$0.0354$*&$0.2031$*&$1.7246$**&$1.2498$&$0.7551$*&$$&$0.0842$**&$0.0809$**&$0.1192$&$ 0.1694$**&$3.0081$**&$1.6163$&$ 0.7686$**&$$&${\bf 0.6839}$\tabularnewline
$\gamma$&$0.3$&$0.1086$**&$0.0574$**&$0.0396$**&$0.2295$**&$1.7305$&$1.2345$**&$0.7221$*&$$&$0.0783$*&$0.0749$*&$0.1099$**&$ 0.1661$*&$3.0264$&$1.5734$**&$ 0.6938$*&$$&$0.6826$\tabularnewline
$\gamma$&$0.5$&$0.1831$&$0.0950$&$0.0638$&$0.3146$&$1.7513$&$1.1887$*&${\bf 0.6080}$&$$&$0.1222$&$0.0960$&$0.1336$&$ 0.2469$&$3.1048$&$1.4748$*&$ {\bf 0.6002}$&$$&$0.6828$*\tabularnewline
\multicolumn{19}{l}{} \\
Mean $=20$&$$&$$&$$&$$&$$&$$&$$&$$&$$&$$&$$&$$&$$&$$&$$&$$&$$&$$\tabularnewline
ML&$$&$2.4694$&$0.6027$&$0.9202$&$1.7694$&$0.4388$&$0.8950$&$2.1682$&$$&$ 6.0983$&$0.4111$&$ 0.9292$&$ 3.1509$&$0.2043$&$0.8152$&$ 4.7388$&$$&$0.4162$\tabularnewline
ML $+$ Cauchit&$$&$0.1015$&$0.9756$&$2.2620$&$1.5232$&$0.1972$&$0.5918$&$3.2111$&$$&$ 3.7708$&$6.7094$&$24.7913$&$ 4.2271$&$0.2695$&$1.2385$&$44.8350$&$$&$0.5779$\tabularnewline
DP&$0.3$&$0.0180$*&${\bf 0.0008}$&${\bf 0.0277}$&$0.0039$*&${\bf 0.0003}$&$0.0069$*&${\bf 0.0036}$&$$&$ {\bf 0.0538}$&${\bf 0.0645}$&$ {\bf 0.1004}$&$ {\bf 0.0948}$&${\bf 0.0284}$&${\bf 0.0417}$&$ {\bf 0.1395}$&$$&$0.6829$\tabularnewline
DP&$0.5$&${\bf 0.0086}$&$0.0056$*&$0.0306$*&${\bf 0.0035}$&$0.0039$*&${\bf 0.0041}$&$0.0175$*&$$&$ 0.0637$**&$0.0722$**&$ 0.1114$**&$ 0.1094$**&$0.0313$**&$0.0469$**&$ 0.1640$**&$$&${\bf 0.6834}$\tabularnewline
$\gamma$&$0.3$&$0.0207$**&$0.0196$**&$0.0392$**&$0.0410$**&$0.0110$**&$0.0178$**&$0.0625$**&$$&$ 0.0566$*&$0.0673$*&$ 0.1047$*&$ 0.1014$*&$0.0297$*&$0.0440$*&$ 0.1508$*&$$&$0.6832$*\tabularnewline
$\gamma$&$0.5$&$0.0802$&$0.0488$&$0.0574$&$0.1059$&$0.0286$&$0.0523$&$0.1524$&$$&$ 0.0794$&$0.0816$&$ 0.1239$&$ 0.1364$&$0.0356$&$0.0558$&$ 0.2112$&$$&$0.6830$**\tabularnewline
\hline
\end{tabular}
}
\end{center}
\label{tb:ne2-probit}
\end{table}
\end{landscape} %

\begin{landscape}
\begin{table}[t]
\caption{Bias, mean square error (MSE), and percentage of correctly classified responses (CCR) using the maximum likelihood, DP and $\gamma$-divergences methods in the ordinal response model with the logit link, and maximum likelihood method with the cauchit link.
The values of the means of the contaminated normal distribution are 5, 10, and 20.
Bold, *, and ** denote the first, second, and third best performance, respectively.}
\begin{center}
\scalebox{0.78}{
\begin{tabular}{rc lllllll c lllllll c l} \hline
& & \multicolumn{7}{l}{Bias} & & \multicolumn{7}{l}{MSE} & & CCR \\ \cline{3-9} \cline{11-17} \cline{19-19}
& tuning & $\beta_1$ & $\beta_2$ & $\beta_3$ & $\delta_1$ & $\delta_2$ & $\delta_3$ & $\delta_4$ & & $\beta_1$ & $\beta_2$ & $\beta_3$ & $\delta_1$ & $\delta_2$ & $\delta_3$ & $\delta_4$ & & \\ \hline 
Mean $=5$&$$&$$&$$&$$&$$&$$&$$&$$&$$&$$&$$&$$&$$&$$&$$&$$&$$&$$\tabularnewline
ML&$$&$1.6318$&$0.2710$&$0.9997$&$0.6608$&${\bf 1.4467}$&$1.5748$&$1.5189$&$$&$2.6980$&$0.1667$&$1.1521$&$0.4995$&${\bf 2.1208}$&$2.5173$&$2.4272$&$$&$0.4579$\tabularnewline
ML $+$ Cauchit&$$&$0.2891$&$0.0649$&$0.4390$&$0.5211$&$1.7503$*&$1.2130$**&${\bf 0.0967}$&$$&$0.2656$&$0.2045$&$0.5773$&$0.5405$&$3.1205$*&$1.5584$**&$0.7093$&$$&$0.5189$\tabularnewline
DP&$0.3$&$0.1896$&${\bf 0.0095}$&$0.1662$**&${\bf 0.2847}$&$1.7596$**&$1.2392$&$0.6453$&$$&$0.1748$&${\bf 0.1350}$&$0.2483$*&${\bf 0.2379}$&$3.1467$**&$1.6020$&$0.6175$&$$&$0.5329$\tabularnewline
DP&$0.5$&${\bf 0.0207}$&$0.0409$**&${\bf 0.0932}$&$0.4532$**&$1.8081$&$1.1690$*&$0.4823$**&$$&${\bf 0.1095}$&$0.1595$**&${\bf 0.2440}$&$0.3710$**&$3.3260$&$1.4431$*&$0.4650$*&$$&${\bf 0.5385}$\tabularnewline
$\gamma$&$0.3$&$0.1439$**&$0.0099$*&$0.1729$&$0.3308$*&$1.7717$&$1.2140$&$0.5869$&$$&$0.1639$**&$0.1400$*&$0.2560$**&$0.2741$*&$3.1916$&$1.5433$&$0.5559$**&$$&$0.5337$**\tabularnewline
$\gamma$&$0.5$&$0.1332$*&$0.0932$&$0.1224$*&$0.5732$&$1.8388$&${\bf 1.1012}$&$0.3248$*&$$&$0.1394$*&$0.1833$&$0.2741$&$0.5178$&$3.4450$&${\bf 1.2996}$&${\bf 0.3729}$&$$&$0.5369$*\tabularnewline
&$$&$$&$$&$$&$$&$$&$$&$$&$$&$$&$$&$$&$$&$$&$$&$$&$$&$$\tabularnewline
Mean $=10$&$$&$$&$$&$$&$$&$$&$$&$$&$$&$$&$$&$$&$$&$$&$$&$$&$$&$$\tabularnewline
ML&$$&$2.2587$&$0.3546$&$1.3786$&$0.8974$&${\bf 1.4037}$&$1.7508$&$1.8827$&$$&$5.1085$&$0.2077$&$2.0074$&$0.8684$&${\bf 2.0003}$&$3.1037$&$3.6562$&$$&$0.3933$\tabularnewline
ML $+$ Cauchit&$$&$0.5227$&${\bf 0.0199}$&$0.5668$&${\bf 0.4481}$&$1.7157$*&$1.2975$&${\bf 0.1614}$&$$&$0.9136$&$0.2468$&$1.0414$&$0.5869$&$3.0172$*&$1.8083$&$1.7474$&$$&$0.5061$\tabularnewline
DP&$0.3$&${\bf 0.0622}$&$0.0346$*&${\bf 0.0441}$&$0.4861$*&$1.8157$**&$1.1586$&$0.4890$&$$&${\bf 0.0859}$&${\bf 0.1441}$&${\bf 0.2095}$&${\bf 0.3833}$&$3.3521$**&$1.4148$&$0.4434$&$$&${\bf 0.5337}$\tabularnewline
DP&$0.5$&$0.0915$*&$0.0466$**&$0.0483$*&$0.5167$**&$1.8229$&$1.1437$&$0.4498$&$$&$0.0991$&$0.1601$&$0.2297$&$0.4326$&$3.3826$&$1.3873$&$0.4364$&$$&${\bf 0.5337}$\tabularnewline
$\gamma$&$0.3$&$0.1100$&$0.0575$&$0.0579$&$0.5379$&$1.8288$&$1.1286$*&$0.4198$**&$$&$0.0979$*&$0.1522$*&$0.2197$*&$0.4443$**&$3.4026$&$1.3499$*&$0.3918$*&$$&$0.5330$\tabularnewline
$\gamma$&$0.5$&$0.2028$&$0.1007$&$0.0828$&$0.6386$&$1.8536$&${\bf 1.0735}$&$0.2869$*&$$&$0.1447$&$0.1854$&$0.2600$&$0.5980$&$3.5027$&${\bf 1.2425}$&${\bf 0.3528}$&$$&${\bf 0.5337}$\tabularnewline
\multicolumn{19}{l}{} \\
Mean $=20$&$$&$$&$$&$$&$$&$$&$$&$$&$$&$$&$$&$$&$$&$$&$$&$$&$$&$$\tabularnewline
ML&$$&$2.4412$&$0.3784$&$1.4846$&$1.3235$&$0.3769$&$0.6316$&$1.4210$&$$&$5.9605$&$0.2196$&$2.2996$&$1.8165$&$0.1727$&$0.4368$&$2.1277$&$$&$0.3819$\tabularnewline
ML $+$ Cauchit&$$&$1.1169$&$0.0661$&$0.9473$&$0.2000$&$0.1815$&$0.3024$&$0.0674$&$$&$2.5090$&$0.8044$&$2.9639$&$0.5394$&$0.1182$&$0.2825$&$1.9291$&$$&$0.4696$\tabularnewline
DP&$0.3$&${\bf 0.0614}$&${\bf 0.0272}$&${\bf 0.0454}$&${\bf 0.0832}$&${\bf 0.0130}$&${\bf 0.0404}$&$0.1050$*&$$&${\bf 0.0778}$&${\bf 0.1424}$&${\bf 0.2146}$&${\bf 0.1431}$&${\bf 0.0543}$&${\bf 0.0723}$&${\bf 0.2073}$&$$&${\bf 0.5417}$\tabularnewline
DP&$0.5$&$0.0795$*&$0.0358$*&$0.0542$*&$0.1039$*&$0.0178$*&$0.0510$*&$0.1338$**&$$&$0.0916$**&$0.1568$**&$0.2357$**&$0.1643$**&$0.0586$**&$0.0801$**&$0.2437$**&$$&$0.5401$\tabularnewline
$\gamma$&$0.3$&$0.1081$**&$0.0498$**&$0.0598$**&$0.1343$**&$0.0258$**&$0.0701$**&$0.1733$&$$&$0.0893$*&$0.1500$*&$0.2252$*&$0.1616$*&$0.0576$*&$0.0793$*&$0.2371$*&$$&$0.5416$*\tabularnewline
$\gamma$&$0.5$&$0.1897$&$0.0891$&$0.0892$&$0.2244$&$0.0479$&$0.1207$&$0.2950$&$$&$0.1336$&$0.1803$&$0.2671$&$0.2265$&$0.0680$&$0.1028$&$0.3477$&$$&$0.5412$**\tabularnewline
\hline
\end{tabular}
}
\end{center}
\label{tb:ne2-logit}
\end{table}
\end{landscape} %

\begin{landscape}
\begin{table}[t]
\caption{Bias, mean square error (MSE), and percentage of correctly classified responses (CCR) using the maximum likelihood, DP and $\gamma$-divergences methods in the ordinal response model with the log-log link, and maximum likelihood method with the cauchit link.
The values of the means of the contaminated normal distribution are 5, 10, and 20.}
\begin{center}
\scalebox{0.78}{
\begin{tabular}{rc lllllll c lllllll c l} \hline
& & \multicolumn{7}{l}{Bias} & & \multicolumn{7}{l}{MSE} & & CCR \\ \cline{3-9} \cline{11-17} \cline{19-19}
& tuning & $\beta_1$ & $\beta_2$ & $\beta_3$ & $\delta_1$ & $\delta_2$ & $\delta_3$ & $\delta_4$ & & $\beta_1$ & $\beta_2$ & $\beta_3$ & $\delta_1$ & $\delta_2$ & $\delta_3$ & $\delta_4$ & & \\ \hline 
Mean $=5$&$$&$$&$$&$$&$$&$$&$$&$$&$$&$$&$$&$$&$$&$$&$$&$$&$$&$$\tabularnewline
ML&$$&$2.2310$&$0.2904$&$0.9532$&$1.9076$&$0.2771$&$1.0800$&$1.6561$&$$&$4.9849$&$0.1253$&$ 0.9820$&$3.6618$&$0.0956$&$1.2086$&$ 2.8527$&$$&$0.5015$\tabularnewline
ML $+$ Cauchit&$$&$1.2603$&$0.6941$&$0.7621$&$2.6900$&$0.1319$&$1.0387$&$2.5270$&$$&$1.9807$&$0.7874$&$ 1.0692$&$7.9889$&$0.1291$&$1.5240$&$ 7.8396$&$$&$0.6733$\tabularnewline
DP&$0.3$&${\bf 0.0791}$&${\bf 0.0483}$&${\bf 0.0323}$&${\bf 0.1102}$&${\bf 0.0311}$&${\bf 0.0784}$&${\bf 0.1407}$&$$&${\bf 0.0867}$&${\bf 0.0866}$&$ {\bf 0.1212}$&${\bf 0.1506}$&${\bf 0.0437}$&${\bf 0.1232}$&$ {\bf 0.2839}$&$$&${\bf 0.6818}$\tabularnewline
DP&$0.5$&$0.1052$*&$0.0610$*&$0.0410$*&$0.1420$*&$0.0407$*&$0.1053$*&$0.1848$*&$$&$0.1082$**&$0.1011$**&$ 0.1388$**&$0.1882$**&$0.0490$**&$0.1492$**&$ 0.3510$**&$$&$0.6813$\tabularnewline
$\gamma$&$0.3$&$0.1304$**&$0.0727$**&$0.0456$**&$0.1725$**&$0.0555$**&$0.1364$**&$0.2281$**&$$&$0.1019$*&$0.0934$*&$ 0.1274$*&$0.1762$*&$0.0480$*&$0.1427$*&$ 0.3322$*&$$&$0.6814$**\tabularnewline
$\gamma$&$0.5$&$0.2286$&$0.1199$&$0.0739$&$0.2942$&$0.0980$&$0.2436$&$0.3948$&$$&$0.1670$&$0.1248$&$ 0.1595$&$0.2864$&$0.0638$&$0.2224$&$ 0.5303$&$$&$0.6815$*\tabularnewline
&$$&$$&$$&$$&$$&$$&$$&$$&$$&$$&$$&$$&$$&$$&$$&$$&$$&$$\tabularnewline
Mean $=10$&$$&$$&$$&$$&$$&$$&$$&$$&$$&$$&$$&$$&$$&$$&$$&$$&$$&$$\tabularnewline
ML&$$&$2.4070$&$0.2550$&$1.0512$&$2.0048$&$0.3002$&$1.1289$&$1.7176$&$$&$5.7949$&$0.1004$**&$ 1.1768$&$4.0342$&$0.1082$&$1.3156$&$ 3.0551$&$$&$0.5142$\tabularnewline
ML $+$ Cauchit&$$&$0.7529$&$0.6222$&$1.3642$&$2.5890$&$0.2144$&$0.7858$&$2.5472$&$$&$2.9669$&$2.0477$&$10.9855$&$8.8889$&$0.2627$&$2.4516$&$19.9793$&$$&$0.6454$\tabularnewline
DP&$0.3$&${\bf 0.0864}$&${\bf 0.0531}$&${\bf 0.0287}$&${\bf 0.1126}$&${\bf 0.0340}$&${\bf 0.0883}$&${\bf 0.1506}$&$$&${\bf 0.0879}$&${\bf 0.0888}$&$ {\bf 0.1178}$&${\bf 0.1489}$&${\bf 0.0449}$&${\bf 0.1250}$&$ {\bf 0.2873}$&$$&${\bf 0.6660}$\tabularnewline
DP&$0.5$&$0.1115$*&$0.0658$*&$0.0384$*&$0.1436$*&$0.0432$*&$0.1144$*&$0.1949$*&$$&$0.1109$**&$0.1027$&$ 0.1357$**&$0.1883$**&$0.0508$**&$0.1519$**&$ 0.3580$**&$$&$0.6657$*\tabularnewline
$\gamma$&$0.3$&$0.1379$**&$0.0777$**&$0.0422$**&$0.1751$**&$0.0585$**&$0.1468$**&$0.2387$**&$$&$0.1041$*&$0.0960$*&$ 0.1239$*&$0.1751$*&$0.0494$*&$0.1458$*&$ 0.3379$*&$$&$0.6657$*\tabularnewline
$\gamma$&$0.5$&$0.2357$&$0.1253$&$0.0717$&$0.2967$&$0.1009$&$0.2538$&$0.4068$&$$&$0.1721$&$0.1272$&$ 0.1559$&$0.2881$&$0.0662$&$0.2284$&$ 0.5445$&$$&$0.6656$\tabularnewline
\multicolumn{19}{l}{} \\
Mean $=20$&$$&$$&$$&$$&$$&$$&$$&$$&$$&$$&$$&$$&$$&$$&$$&$$&$$&$$\tabularnewline
ML&$$&$2.4717$&$0.2387$&$1.0820$&$2.0293$&$0.3192$&$1.1554$&$1.7542$&$$&$6.1098$&$0.0903$&$ 1.2414$&$ 4.1320$&$0.1209$&$1.3758$&$ 3.1793$&$$&$0.5044$\tabularnewline
ML $+$ Cauchit&$$&$0.8895$&$0.1734$&$2.6148$&$1.8459$&$0.5027$&$0.2019$&$1.5774$&$$&$4.1309$&$2.3968$&$29.9653$&$ 5.4273$&$0.4916$&$2.3754$&$28.1556$&$$&$0.5742$\tabularnewline
DP&$0.3$&${\bf 0.0649}$&${\bf 0.0439}$&${\bf 0.0296}$&${\bf 0.0951}$&${\bf 0.0271}$&${\bf 0.0665}$&${\bf 0.1138}$&$$&${\bf 0.0802}$&${\bf 0.0885}$&$ {\bf 0.1181}$&$ {\bf 0.1427}$&${\bf 0.0430}$&${\bf 0.1159}$&$ {\bf 0.2642}$&$$&$0.6684$*\tabularnewline
DP&$0.5$&$0.0877$*&$0.0555$*&$0.0377$*&$0.1225$*&$0.0358$*&$0.0906$*&$0.1542$*&$$&$0.1002$**&$0.1019$**&$ 0.1344$**&$ 0.1768$**&$0.0485$**&$0.1402$**&$ 0.3257$**&$$&$0.6682$**\tabularnewline
$\gamma$&$0.3$&$0.1159$*&$0.0683$*&$0.0431$*&$0.1573$*&$0.0516$*&$0.1245$*&$0.2012$*&$$&$0.0940$*&$0.0953$*&$ 0.1242$*&$ 0.1666$*&$0.0472$*&$0.1339$*&$ 0.3075$*&$$&$0.6681$\tabularnewline
$\gamma$&$0.5$&$0.2105$&$0.1145$&$0.0704$&$0.2741$&$0.0936$&$0.2286$&$0.3634$&$$&$0.1542$&$0.1249$&$ 0.1542$&$ 0.2682$&$0.0630$&$0.2087$&$ 0.4903$&$$&${\bf 0.6693}$\tabularnewline
\hline
\end{tabular}
}
\end{center}
\label{tb:ne2-loglog}
\end{table}
\end{landscape} %

\subsection{Real data analysis}
This section shows the robustness and usefulness of the proposed robust ordinal response with the DP and $\gamma$-divergences through the analysis of two real datasets: Boston housing data (\citealp{harrison1978hedonic}) and Affairs data (\citealp{fair1978theory}).

The Boston housing data is often used to evaluate the performance of robust inference methods in models with the continuous data as the response variable, such as the linear regression (\citealp{hashimoto2020robust, hamura2022log, kawakami2022approximate}), but in this case, the response variable, the corrected median value of owner-occupied homes in USD 1000's is applied to the ordinal response model assuming that it is obtained as the ordered categorical data with five categories.
The Boston housing data has 14 variables including one binary variable.
The Affairs data contains 3 continuous, 2 binary, 2 ordered categorical, and 2 multinomial categorical variables.
We set one of the ordered categorical data, the self rating of marriage as the response variable.
In the two real data, the continuous data are standardized to have mean 0 and variance 1, the ordered categorical data in the covariates is numerically transformed using the Likert sigma method, and the multinomial categorical data are transformed using the dummy variables.
In analyzing the two sets of real data, we use the commonly-used symmetric link functions, the probit and logit links.



\begin{figure}[H]
\begin{center}
\includegraphics[scale=0.5]{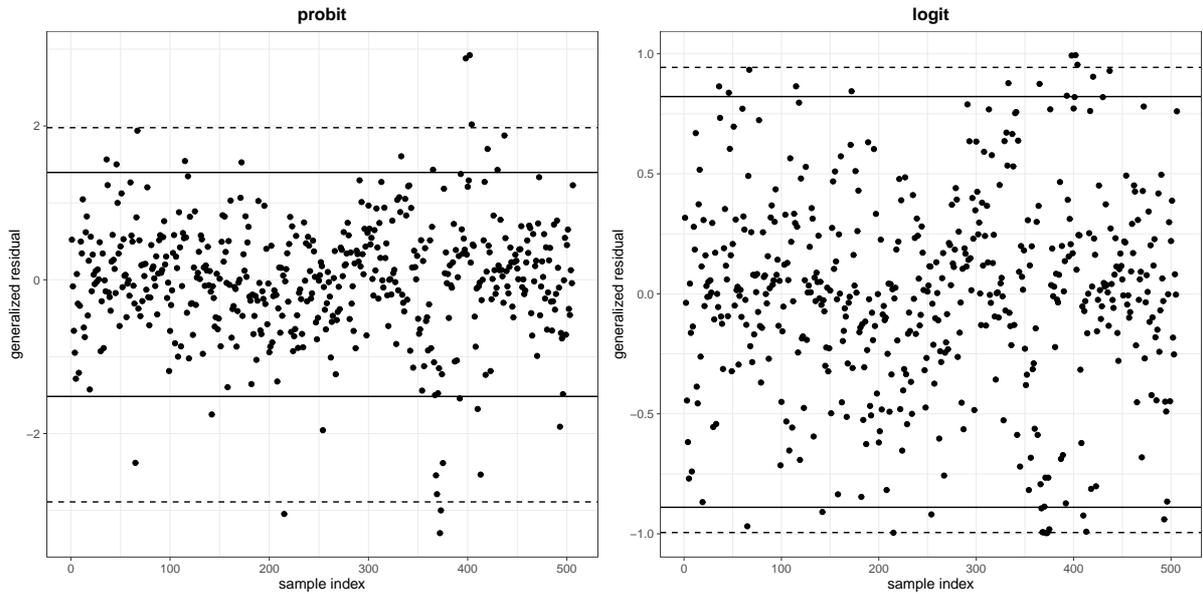}
\caption{The generalized residuals (\citealp{franses2001quantitative}) plots for the Boston housing data with the probit and logit links.
The solid and dashed lines indicate 95\% and 99\% intervals of the generalized residuals, respectively.}
\label{fg:grd_boston}
\end{center}
\end{figure}

\begin{figure}[H]
\begin{center}
\includegraphics[scale=0.5]{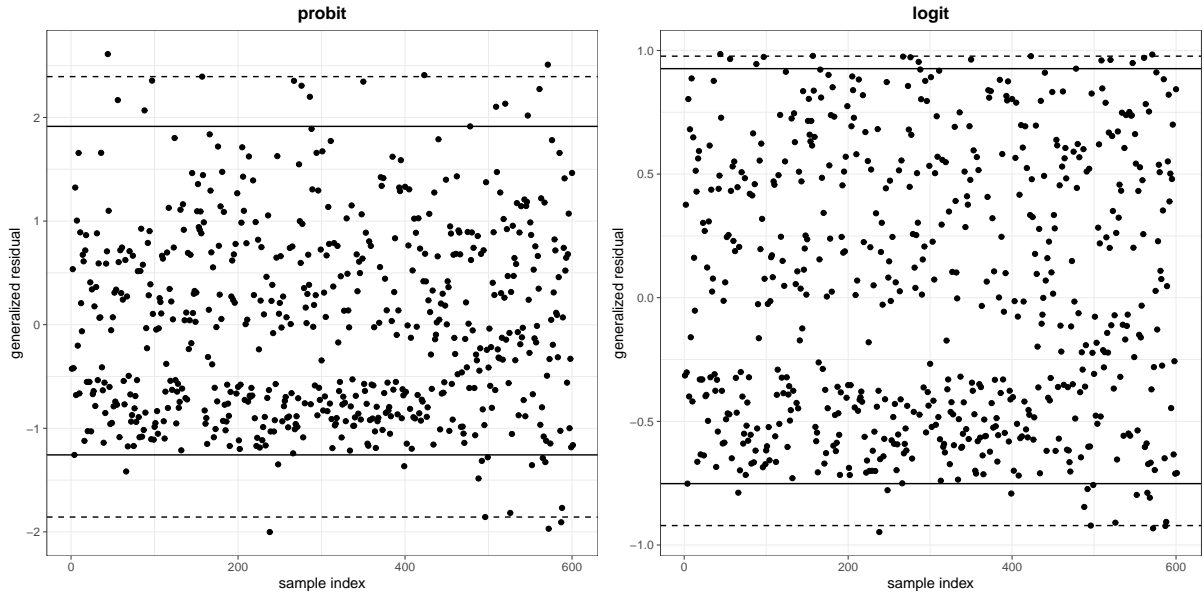}
\caption{The generalized residuals (\citealp{franses2001quantitative}) plots for the Affairs data with the probit and logit links.
Solid and dashed lines indicate 95\% and 99\% intervals of the generalized residuals, respectively.}
\label{fg:grd_affairs}
\end{center}
\end{figure}

The generalized residuals (\citealp{franses2001quantitative}) plots with two link functions for two real datasets are shown in Figures \ref{fg:grd_boston} and \ref{fg:grd_affairs}.
The solid and dashed lines in these figures indicate 95\% and 99\% intervals of the generalized residuals, respectively.
In Subsections \ref{sec:boston} and \ref{sec:affairs}, the original data and the modified data in which individuals with the values of the generalized residuals larger than the 95\% interval (solid line) in Figures \ref{fg:grd_boston} and \ref{fg:grd_affairs} are removed are used for the data analyses.

\subsubsection{Boston housing data}
\label{sec:boston}
In this section, we consider the Boston data (\citealp{harrison1978hedonic}) that contains 506 individuals and 14 variables including one binary variable.
As mentioned above, the Boston housing data is often used in the numerical experiments of the robust inference methods for the continuous data as the response variable (\citealp{hashimoto2020robust, hamura2022log, kawakami2022approximate}).
Hence, we transform the continuous response variable, the corrected median value of owner-occupied homes in USD 1000's to the the ordered categorical response variable with five categories ($\leq10, 10< x \leq 20, 20< x \leq 30, 30< x \leq 40, 40<$).
We also standardize 12 continuous valued covariates as mean 0 and variance 1.
For the link function in the data analysis, we use the commonly-used symmetric link functions, the probit and logit links.

Table \ref{tb:result_boston} shows the estimates of each parameter when applying the maximum likelihood method and the proposed robust ordinal response model with the DP and $\gamma$-divergences ($\alpha, \gamma = 0.3, 0.5$) for the probit and logit links to the original Boston housing data and the modified data excluding individuals whose values of the generalized residuals are larger than the 95\% interval (Figure \ref{fg:grd_boston}).
The values in the row of ``Distance" in the column of each method are on the left, the sum of squares of the differences between the estimated values of each parameter applying the ML to the modified data and the corresponding method to the original data divided by the number of parameters, and on the right, that between the estimated values of each parameter applying the corresponding method to the original and modified data divided by the number of parameters.
Since these two values are the same in the ML, they are listed only in the ``original'' column.
The smaller these two values are, the less sensitive the inference method is to outliers.

It can be seen from Table \ref{tb:result_boston} that the values of each ``Distance'' are smaller for all of our proposed methods, both with the probit and logit links, compared to ML.
Namely, our proposed methods with the robust divergences are able to eliminate the influence of data that appear to be outliers.
Furthermore, for the same tuning parameter values, the inferences with the $\gamma$-divergence have smaller values of the "Distance" than those with the DP-divergence.
The reason why the values of ``Distance'' are not closer to zero may be that it is difficult to identify outliers by using the generalized residuals for high-dimensional data, and therefore, the data whose values of the generalized residuals are larger than the 95\% interval do not always coincide with the data whose influence on the inference is removed by the proposed methods.


\begin{table}[H]
\caption{Estimates of each parameter when applying the maximum likelihood method and the proposed robust ordinal response model with the DP and $\gamma$-divergences ($\alpha, \gamma = 0.3, 0.5$) for the probit and logit links to the original Boston housing data and the modified data excluding individuals whose values of the generalized residuals are larger than the 95\% interval (Figure \ref{fg:grd_boston}).
The values in the row of ``Distance" in the column of each method are on the left, the sum of squares of the differences between the estimated values of each parameter applying the ML to the modified data and the corresponding method to the original data divided by the number of parameters, and on the right, that between the estimated values of each parameter applying the corresponding method to the original and modified data divided by the number of parameters.}
\begin{center}
\scalebox{0.70}{
\begin{tabular}{c} 
\begin{tabular}{l rr c rr c rr c rr c rr} \hline
 & \multicolumn{2}{c}{ML} & & \multicolumn{2}{c}{DP-div.: $\alpha=0.3$} & & \multicolumn{2}{c}{DP-div.: $\alpha=0.5$} & & \multicolumn{2}{c}{$\gamma$-div.: $\gamma=0.3$} & & \multicolumn{2}{c}{$\gamma$-div.: $\gamma=0.5$} \\ \cline{2-3} \cline{5-6} \cline{8-9} \cline{11-12} \cline{14-15}
probit & original & modified & & original & modified & & original & modified & & original & modified & & original & modified \\ \hline
Coefficient & \multicolumn{14}{c}{} \\
$\beta_{1}$: crim&$-0.3745$&$-0.6241$&$$&$-0.4306$&$-0.5983$&$$&$-0.4545$&$-0.5867$&$$&$-0.4333$&$-0.5986$&$$&$-0.4664$&$-0.5891$\tabularnewline
$\beta_{2}$: zn&$ 0.2123$&$ 0.3280$&$$&$ 0.2235$&$ 0.3314$&$$&$ 0.2340$&$ 0.3372$&$$&$ 0.2244$&$ 0.3315$&$$&$ 0.2386$&$ 0.3385$\tabularnewline
$\beta_{3}$: indus&$ 0.2172$&$ 0.2416$&$$&$ 0.2471$&$ 0.2478$&$$&$ 0.2704$&$ 0.2626$&$$&$ 0.2486$&$ 0.2480$&$$&$ 0.2773$&$ 0.2639$\tabularnewline
$\beta_{4}$: chas&$ 0.5537$&$ 0.7009$&$$&$ 0.4574$&$ 0.6052$&$$&$ 0.4737$&$ 0.5693$&$$&$ 0.4595$&$ 0.6054$&$$&$ 0.4853$&$ 0.5716$\tabularnewline
$\beta_{5}$: nox&$-0.5457$&$-0.5447$&$$&$-0.4104$&$-0.5073$&$$&$-0.3659$&$-0.4757$&$$&$-0.4116$&$-0.5076$&$$&$-0.3699$&$-0.4771$\tabularnewline
$\beta_{6}$: rm&$ 0.4454$&$ 0.8548$&$$&$ 0.8347$&$ 0.9461$&$$&$ 1.0661$&$ 1.1446$&$$&$ 0.8411$&$ 0.9468$&$$&$ 1.0994$&$ 1.1526$\tabularnewline
$\beta_{7}$: age&$-0.1257$&$-0.1802$&$$&$-0.2200$&$-0.1903$&$$&$-0.2838$&$-0.2643$&$$&$-0.2214$&$-0.1904$&$$&$-0.2920$&$-0.2665$\tabularnewline
$\beta_{8}$: dis&$-0.6845$&$-0.9430$&$$&$-0.6540$&$-0.8803$&$$&$-0.6483$&$-0.8655$&$$&$-0.6566$&$-0.8806$&$$&$-0.6592$&$-0.8687$\tabularnewline
$\beta_{9}$: rad&$ 0.9869$&$ 1.5333$&$$&$ 1.0109$&$ 1.4507$&$$&$ 1.0578$&$ 1.4118$&$$&$ 1.0161$&$ 1.4514$&$$&$ 1.0840$&$ 1.4172$\tabularnewline
$\beta_{10}$: tax&$-0.7900$&$-1.2633$&$$&$-0.9249$&$-1.2366$&$$&$-1.0285$&$-1.2759$&$$&$-0.9302$&$-1.2372$&$$&$-1.0556$&$-1.2817$\tabularnewline
$\beta_{11}$: ptratio&$-0.5598$&$-0.8484$&$$&$-0.6101$&$-0.8033$&$$&$-0.6435$&$-0.8087$&$$&$-0.6134$&$-0.8038$&$$&$-0.6584$&$-0.8124$\tabularnewline
$\beta_{12}$: black&$ 0.3314$&$ 0.6214$&$$&$ 0.5721$&$ 0.6752$&$$&$ 0.6814$&$ 0.7490$&$$&$ 0.5765$&$ 0.6755$&$$&$ 0.7031$&$ 0.7535$\tabularnewline
$\beta_{13}$: lstat&$-1.3334$&$-2.2568$&$$&$-1.4068$&$-2.1002$&$$&$-1.4452$&$-1.9575$&$$&$-1.4147$&$-2.1012$&$$&$-1.4797$&$-1.9641$\tabularnewline
Distance &$0.1445$&\multicolumn{1}{c}{$-$}&$$&$ 0.1062$&$ 0.0734$&$$&$ 0.0961$&$ 0.0449$&$$&$ 0.1040$&$ 0.0717$&$$&$ 0.0889$&$ 0.0396$\tabularnewline
\multicolumn{15}{c}{} \\
Cutoff & \multicolumn{14}{c}{} \\
$\delta_1$&$-4.2923$&$-7.2768$&$$&$-5.2172$&$-7.1307$&$$&$-5.8103$&$-7.2795$&$$&$-5.2502$&$-7.1343$&$$&$-5.9749$&$-7.3150$\tabularnewline
$\delta_2$&$-0.3806$&$-0.5431$&$$&$-0.3763$&$-0.5100$&$$&$-0.3756$&$-0.5242$&$$&$-0.3777$&$-0.5102$&$$&$-0.3811$&$-0.5265$\tabularnewline
$\delta_3$&$ 2.5033$&$ 4.0394$&$$&$ 3.0219$&$ 3.9679$&$$&$ 3.3426$&$ 4.0879$&$$&$ 3.0410$&$ 3.9700$&$$&$ 3.4325$&$ 4.1083$\tabularnewline
$\delta_4$&$ 3.7920$&$ 6.1554$&$$&$ 4.7409$&$ 6.0992$&$$&$ 5.3043$&$ 6.3753$&$$&$ 4.7685$&$ 6.1023$&$$&$ 5.4377$&$ 6.4065$\tabularnewline
Distance&$4.2197$&\multicolumn{1}{c}{$-$}&$$&$ 1.8265$&$ 1.6048$&$$&$ 0.8472$&$ 0.9708$&$$&$ 1.7638$&$ 1.5523$&$$&$ 0.6512$&$ 0.8031$\tabularnewline
\hline
\end{tabular}\\
\\
\begin{tabular}{l rr c rr c rr c rr c rr} \hline
 & \multicolumn{2}{c}{ML} & & \multicolumn{2}{c}{DP-div.: $\alpha=0.3$} & & \multicolumn{2}{c}{DP-div.: $\alpha=0.5$} & & \multicolumn{2}{c}{$\gamma$-div.: $\gamma=0.3$} & & \multicolumn{2}{c}{$\gamma$-div.: $\gamma=0.5$} \\ \cline{2-3} \cline{5-6} \cline{8-9} \cline{11-12} \cline{14-15}
logit & original & modified & & original & modified & & original & modified & & original & modified & & original & modified \\ \hline
Coefficient & \multicolumn{14}{c}{} \\
$\beta_{1}$: crim&$-0.7092$&$-1.1042$&$$&$-0.7492$&$-1.0500$&$$&$-0.7689$&$-1.0483$&$$&$-0.7529$&$-1.0502$&$$&$-0.7874$&$-1.0516$\tabularnewline
$\beta_{2}$: zn&$ 0.3921$&$ 0.5884$&$$&$ 0.3936$&$ 0.5871$&$$&$ 0.4087$&$ 0.6022$&$$&$ 0.3951$&$ 0.5871$&$$&$ 0.4168$&$ 0.6040$\tabularnewline
$\beta_{3}$: indus&$ 0.3760$&$ 0.4501$&$$&$ 0.4157$&$ 0.4541$&$$&$ 0.4460$&$ 0.4658$&$$&$ 0.4178$&$ 0.4543$&$$&$ 0.4563$&$ 0.4672$\tabularnewline
$\beta_{4}$: chas&$ 0.8528$&$ 1.1848$&$$&$ 0.7813$&$ 1.0213$&$$&$ 0.8059$&$ 0.9717$&$$&$ 0.7849$&$ 1.0214$&$$&$ 0.8240$&$ 0.9745$\tabularnewline
$\beta_{5}$: nox&$-0.8716$&$-0.9267$&$$&$-0.6739$&$-0.8367$&$$&$-0.6023$&$-0.7828$&$$&$-0.6758$&$-0.8368$&$$&$-0.6088$&$-0.7844$\tabularnewline
$\beta_{6}$: rm&$ 0.9748$&$ 1.6316$&$$&$ 1.5257$&$ 1.8333$&$$&$ 1.8650$&$ 2.0513$&$$&$ 1.5352$&$ 1.8339$&$$&$ 1.9169$&$ 2.0594$\tabularnewline
$\beta_{7}$: age&$-0.2244$&$-0.3350$&$$&$-0.3951$&$-0.3906$&$$&$-0.5077$&$-0.4721$&$$&$-0.3974$&$-0.3908$&$$&$-0.5211$&$-0.4739$\tabularnewline
$\beta_{8}$: dis&$-1.1820$&$-1.6192$&$$&$-1.1127$&$-1.5037$&$$&$-1.1010$&$-1.4781$&$$&$-1.1167$&$-1.5039$&$$&$-1.1188$&$-1.4819$\tabularnewline
$\beta_{9}$: rad&$ 1.7719$&$ 2.6890$&$$&$ 1.7614$&$ 2.5267$&$$&$ 1.8128$&$ 2.5041$&$$&$ 1.7691$&$ 2.5274$&$$&$ 1.8537$&$ 2.5113$\tabularnewline
$\beta_{10}$: tax&$-1.4505$&$-2.2116$&$$&$-1.6220$&$-2.1835$&$$&$-1.7665$&$-2.2500$&$$&$-1.6297$&$-2.1843$&$$&$-1.8079$&$-2.2570$\tabularnewline
$\beta_{11}$: ptratio&$-1.0137$&$-1.4820$&$$&$-1.0525$&$-1.4037$&$$&$-1.0947$&$-1.4056$&$$&$-1.0573$&$-1.4041$&$$&$-1.1175$&$-1.4098$\tabularnewline
$\beta_{12}$: black&$ 0.7336$&$ 1.3913$&$$&$ 1.0345$&$ 1.4556$&$$&$ 1.1608$&$ 1.5338$&$$&$ 1.0407$&$ 1.4560$&$$&$ 1.1928$&$ 1.5393$\tabularnewline
$\beta_{13}$: lstat&$-2.4741$&$-3.8863$&$$&$-2.4362$&$-3.5245$&$$&$-2.4360$&$-3.3565$&$$&$-2.4471$&$-3.5251$&$$&$-2.4874$&$-3.3653$\tabularnewline
Distance&$0.3857$&\multicolumn{1}{c}{$-$}&$$&$ 0.3297$&$ 0.2190$&$$&$ 0.3091$&$ 0.1653$&$$&$ 0.3238$&$ 0.2142$&$$&$ 0.2860$&$ 0.1479$\tabularnewline
\multicolumn{15}{c}{} \\
Cutoff & \multicolumn{14}{c}{} \\
$\delta_1$&$-8.2184$&$-13.1173$&$$&$-9.2307$&$-12.8034$&$$&$-9.9826$&$-13.0503$&$$&$-9.2789$&$-12.8064$&$$&$-10.2354$&$-13.0931$\tabularnewline
$\delta_2$&$-0.6643$&$ -0.8915$&$$&$-0.6450$&$ -0.8540$&$$&$-0.6628$&$ -0.8716$&$$&$-0.6471$&$ -0.8542$&$$&$ -0.6733$&$ -0.8742$\tabularnewline
$\delta_3$&$ 4.7262$&$  7.2210$&$$&$ 5.3122$&$  7.0599$&$$&$ 5.7264$&$  7.1793$&$$&$ 5.3398$&$  7.0616$&$$&$  5.8638$&$  7.2024$\tabularnewline
$\delta_4$&$ 7.2010$&$ 11.0035$&$$&$ 8.3303$&$ 10.8968$&$$&$ 9.0904$&$ 11.1933$&$$&$ 8.3706$&$ 10.8994$&$$&$  9.2958$&$ 11.2281$\tabularnewline
Distance&$11.1833$&\multicolumn{1}{c}{$-$}&$$&$ 6.4889$&$  5.6122$&$$&$ 3.9431$&$  3.9969$&$$&$ 6.3161$&$  5.4615$&$$&$  3.2777$&$  3.4331$\tabularnewline
\hline
\end{tabular}
\end{tabular}
}
\end{center}
\label{tb:result_boston}
\end{table}%

\subsubsection{Affairs data}
\label{sec:affairs}
Consider the Affairs data (\citealp{fair1978theory}) that contains 601 observations and 9 variables including 3 continuous, 2 binary, 2 ordered categorical, and 2 multinomial categorical variables.
We analyze the data with the self rating of marriage as the ordered categorical response variable, and then we standardize 3 continuous valued covariates as mean 0 and variance 1, transform one ordered categorical covariate by the Likert sigma method, and create dummy variables from the multinomial categorical covariates.
For the link function in the data analysis, we use the commonly-used symmetric link functions, the probit and logit links.

Table \ref{tb:result_affairs} shows the estimates of each parameter when applying the maximum likelihood method and the proposed robust ordinal response model with the DP and $\gamma$-divergences ($\alpha, \gamma = 0.3, 0.5$) for the probit and logit links to the original Affairs data and the modified data excluding individuals whose values of the generalized residuals are larger than the 95\% interval (Figure \ref{fg:grd_affairs}).
The values in the row of ``Distance" in the column of each method are on the left, the sum of squares of the differences between the estimated values of each parameter applying the ML to the modified data and the corresponding method to the original data divided by the number of parameters, and on the right, that between the estimated values of each parameter applying the corresponding method to the original and modified data divided by the number of parameters.
Since these two values are the same in the ML, they are listed only in the ``original'' column.
The smaller these two values are, the less sensitive the inference method is to outliers.

It can be seen from Table \ref{tb:result_affairs} that the values of each ``Distance'' are smaller for all of our proposed methods, with both the probit and logit links, compared to ML.
Namely, our proposed methods with the robust divergences are able to eliminate the influence of data that appear to be outliers.
The values of "Distance" in our proposed inference methods with the DP and $\gamma$-divergences are about the same.
The reason why the values of ``Distance'' are not closer to zero may be that it is difficult to identify outliers by using the generalized residuals for high-dimensional data, and therefore, the data whose values of the generalized residuals are larger than the 95\% interval do not always coincide with the data whose influence on the inference is removed by the proposed methods.


\begin{table}[H]
\caption{Estimates of each parameter when applying the maximum likelihood method and the proposed robust ordinal response model with the DP and $\gamma$-divergences ($\alpha, \gamma = 0.3, 0.5$) for the probit and logit links to the original Affairs data and the modified data excluding individuals whose values of the generalized residuals are larger than the 95\% interval (Figure \ref{fg:grd_affairs}).
The values in the row of ``Distance" in the column of each method are on the left, the sum of squares of the differences between the estimated values of each parameter applying the ML to the modified data and the corresponding method to the original data divided by the number of parameters, and on the right, that between the estimated values of each parameter applying the corresponding method to the original and modified data divided by the number of parameters.}
\begin{center}
\scalebox{0.7}{
\begin{tabular}{c}
\begin{tabular}{l rr c rr c rr c rr c rr} \hline
 & \multicolumn{2}{c}{ML} & & \multicolumn{2}{c}{DP-div.: $\alpha=0.3$} & & \multicolumn{2}{c}{DP-div.: $\alpha=0.5$} & & \multicolumn{2}{c}{$\gamma$-div.: $\gamma=0.3$} & & \multicolumn{2}{c}{$\gamma$-div.: $\gamma=0.5$} \\ \cline{2-3} \cline{5-6} \cline{8-9} \cline{11-12} \cline{14-15}
probit & original & modified & & original & modified & & original & modified & & original & modified & & original & modified \\ \hline
Coefficient & \multicolumn{14}{c}{} \\
$\beta_{1}$: affairs&$-0.2273$&$-0.4026$&$$&$-0.2465$&$-0.3898$&$$&$-0.2616$&$-0.3844$&$$&$-0.2466$&$-0.3897$&$$&$-0.2625$&$-0.3845$\tabularnewline
$\beta_{2}$: gender&$-0.0631$&$-0.0432$&$$&$-0.0826$&$-0.0575$&$$&$-0.0886$&$-0.0674$&$$&$-0.0826$&$-0.0575$&$$&$-0.0888$&$-0.0673$\tabularnewline
$\beta_{3}$: age&$-0.0660$&$-0.1369$&$$&$-0.0628$&$-0.1280$&$$&$-0.0660$&$-0.1253$&$$&$-0.0629$&$-0.1281$&$$&$-0.0662$&$-0.1253$\tabularnewline
$\beta_{4}$: yearsmarried&$-0.1281$&$-0.0745$&$$&$-0.1327$&$-0.0979$&$$&$-0.1343$&$-0.1103$&$$&$-0.1327$&$-0.0979$&$$&$-0.1346$&$-0.1103$\tabularnewline
$\beta_{5}$: children&$-0.2730$&$-0.3264$&$$&$-0.2863$&$-0.3091$&$$&$-0.2956$&$-0.3049$&$$&$-0.2864$&$-0.3091$&$$&$-0.2965$&$-0.3050$\tabularnewline
$\beta_{6}$: religiousness&$ 0.0451$&$ 0.0313$&$$&$ 0.0348$&$ 0.0231$&$$&$ 0.0304$&$ 0.0201$&$$&$ 0.0348$&$ 0.0231$&$$&$ 0.0304$&$ 0.0201$\tabularnewline
$\beta_{7}\sim\beta_{12}$: &$ 0.7429$&$ 1.5177$&$$&$ 0.8703$&$ 1.5393$&$$&$ 1.0083$&$ 1.6442$&$$&$ 0.8711$&$ 1.5412$&$$&$ 1.0065$&$ 1.6437$\tabularnewline
&$ 1.1135$&$ 1.9891$&$$&$ 1.2416$&$ 1.9785$&$$&$ 1.3809$&$ 2.0636$&$$&$ 1.2425$&$ 1.9803$&$$&$ 1.3800$&$ 2.0632$\tabularnewline
education&$ 1.3070$&$ 2.1821$&$$&$ 1.4236$&$ 2.1566$&$$&$ 1.5525$&$ 2.2302$&$$&$ 1.4246$&$ 2.1585$&$$&$ 1.5519$&$ 2.2297$\tabularnewline
&$ 1.1534$&$ 2.1617$&$$&$ 1.3418$&$ 2.1820$&$$&$ 1.5145$&$ 2.2800$&$$&$ 1.3428$&$ 2.1839$&$$&$ 1.5144$&$ 2.2796$\tabularnewline
&$ 1.0974$&$ 1.9841$&$$&$ 1.2550$&$ 1.9870$&$$&$ 1.4066$&$ 2.0789$&$$&$ 1.2559$&$ 1.9889$&$$&$ 1.4058$&$ 2.0785$\tabularnewline
&$ 1.3757$&$ 2.3220$&$$&$ 1.5158$&$ 2.3090$&$$&$ 1.6626$&$ 2.3935$&$$&$ 1.5168$&$ 2.3109$&$$&$ 1.6625$&$ 2.3930$\tabularnewline
$\beta_{13}\sim\beta_{18}$: &$ 0.3589$&$ 0.4829$&$$&$ 0.2806$&$ 0.3212$&$$&$ 0.2405$&$ 0.2426$&$$&$ 0.2808$&$ 0.3215$&$$&$ 0.2398$&$ 0.2425$\tabularnewline
&$-0.2767$&$-0.3298$&$$&$-0.2390$&$-0.3097$&$$&$-0.2198$&$-0.2969$&$$&$-0.2390$&$-0.3096$&$$&$-0.2202$&$-0.2970$\tabularnewline
occupation&$ 0.0559$&$ 0.0475$&$$&$ 0.0591$&$ 0.0393$&$$&$ 0.0590$&$ 0.0394$&$$&$ 0.0592$&$ 0.0394$&$$&$ 0.0589$&$ 0.0393$\tabularnewline
&$ 0.0621$&$ 0.0287$&$$&$ 0.0352$&$ 0.0026$&$$&$ 0.0170$&$-0.0079$&$$&$ 0.0352$&$ 0.0027$&$$&$ 0.0168$&$-0.0079$\tabularnewline
&$-0.0187$&$-0.0591$&$$&$-0.0472$&$-0.0839$&$$&$-0.0683$&$-0.0967$&$$&$-0.0472$&$-0.0838$&$$&$-0.0687$&$-0.0967$\tabularnewline
&$-0.7293$&$-0.8639$&$$&$-0.7464$&$-0.8580$&$$&$-0.7664$&$-0.8550$&$$&$-0.7466$&$-0.8578$&$$&$-0.7686$&$-0.8551$\tabularnewline
Distance&$0.2729$&\multicolumn{1}{c}{$-$}&$$&$ 0.1948$&$ 0.1916$&$$&$ 0.1292$&$ 0.1632$&$$&$ 0.1943$&$ 0.1920$&$$&$ 0.1294$&$ 0.1633$\tabularnewline
\multicolumn{15}{c}{} \\
Cutoff & \multicolumn{14}{c}{} \\
$\delta_1$&$-1.2631$&$-1.5275$&$$&$-1.2035$&$-1.2956$&$$&$-1.1385$&$-1.1313$&$$&$-1.2034$&$-1.2937$&$$&$-1.1453$&$-1.1322$\tabularnewline
$\delta_2$&$-0.2932$&$ 0.4043$&$$&$-0.1876$&$ 0.4033$&$$&$-0.0719$&$ 0.4874$&$$&$-0.1871$&$ 0.4053$&$$&$-0.0762$&$ 0.4867$\tabularnewline
$\delta_3$&$ 0.3226$&$ 1.1351$&$$&$ 0.4195$&$ 1.1032$&$$&$ 0.5327$&$ 1.1700$&$$&$ 0.4201$&$ 1.1052$&$$&$ 0.5286$&$ 1.1693$\tabularnewline
$\delta_4$&$ 1.2436$&$ 2.1633$&$$&$ 1.3508$&$ 2.1269$&$$&$ 1.4745$&$ 2.1948$&$$&$ 1.3515$&$ 2.1288$&$$&$ 1.4718$&$ 2.1942$\tabularnewline
Distance&$0.5156$&\multicolumn{1}{c}{$-$}&$$&$ 0.4069$&$ 0.3568$&$$&$ 0.3039$&$ 0.3095$&$$&$ 0.4062$&$ 0.3582$&$$&$ 0.3057$&$ 0.3123$\tabularnewline
\hline
\end{tabular}\\
\\
\begin{tabular}{l rr c rr c rr c rr c rr} \hline
 & \multicolumn{2}{c}{ML} & & \multicolumn{2}{c}{DP-div.: $\alpha=0.3$} & & \multicolumn{2}{c}{DP-div.: $\alpha=0.5$} & & \multicolumn{2}{c}{$\gamma$-div.: $\gamma=0.3$} & & \multicolumn{2}{c}{$\gamma$-div.: $\gamma=0.5$} \\ \cline{2-3} \cline{5-6} \cline{8-9} \cline{11-12} \cline{14-15}
logit & original & modified & & original & modified & & original & modified & & original & modified & & original & modified \\ \hline
Coefficient & \multicolumn{14}{c}{} \\
$\beta_{1}$: affairs&$-0.4181$&$-0.6741$&$$&$-0.4383$&$-0.6423$&$$&$-0.4521$&$-0.6271$&$$&$-0.4384$&$-0.6421$&$$&$-0.4530$&$-0.6269$\tabularnewline
$\beta_{2}$: gender&$-0.1424$&$-0.1107$&$$&$-0.1623$&$-0.1341$&$$&$-0.1684$&$-0.1478$&$$&$-0.1623$&$-0.1340$&$$&$-0.1688$&$-0.1477$\tabularnewline
$\beta_{3}$: age&$-0.1077$&$-0.2182$&$$&$-0.1064$&$-0.2051$&$$&$-0.1125$&$-0.2002$&$$&$-0.1064$&$-0.2051$&$$&$-0.1127$&$-0.2001$\tabularnewline
$\beta_{4}$: yearsmarried&$-0.2196$&$-0.1568$&$$&$-0.2281$&$-0.1885$&$$&$-0.2318$&$-0.2042$&$$&$-0.2282$&$-0.1884$&$$&$-0.2323$&$-0.2041$\tabularnewline
$\beta_{5}$: children&$-0.4385$&$-0.4861$&$$&$-0.4513$&$-0.4668$&$$&$-0.4604$&$-0.4651$&$$&$-0.4514$&$-0.4667$&$$&$-0.4611$&$-0.4651$\tabularnewline
$\beta_{6}$: religiousness&$ 0.0558$&$ 0.0339$&$$&$ 0.0451$&$ 0.0285$&$$&$ 0.0413$&$ 0.0280$&$$&$ 0.0451$&$ 0.0285$&$$&$ 0.0414$&$ 0.0280$\tabularnewline
$\beta_{7}\sim\beta_{12}$: &$ 1.4611$&$ 2.9214$&$$&$ 1.7035$&$ 3.1848$&$$&$ 1.9450$&$ 3.4666$&$$&$ 1.7071$&$ 3.1840$&$$&$ 1.9520$&$ 3.4751$\tabularnewline
&$ 2.0752$&$ 3.6835$&$$&$ 2.3010$&$ 3.8698$&$$&$ 2.5294$&$ 4.1064$&$$&$ 2.3048$&$ 3.8689$&$$&$ 2.5373$&$ 4.1148$\tabularnewline
education&$ 2.4213$&$ 4.0196$&$$&$ 2.6170$&$ 4.1745$&$$&$ 2.8237$&$ 4.3909$&$$&$ 2.6210$&$ 4.1735$&$$&$ 2.8320$&$ 4.3994$\tabularnewline
&$ 2.2395$&$ 4.0303$&$$&$ 2.5479$&$ 4.2513$&$$&$ 2.8119$&$ 4.4993$&$$&$ 2.5520$&$ 4.2503$&$$&$ 2.8209$&$ 4.5078$\tabularnewline
&$ 2.1284$&$ 3.7342$&$$&$ 2.3874$&$ 3.9352$&$$&$ 2.6267$&$ 4.1795$&$$&$ 2.3914$&$ 3.9343$&$$&$ 2.6349$&$ 4.1881$\tabularnewline
&$ 2.5847$&$ 4.2791$&$$&$ 2.8177$&$ 4.4505$&$$&$ 3.0479$&$ 4.6804$&$$&$ 2.8217$&$ 4.4494$&$$&$ 3.0568$&$ 4.6888$\tabularnewline
$\beta_{13}\sim\beta_{18}$: &$ 0.5475$&$ 0.6673$&$$&$ 0.4128$&$ 0.4197$&$$&$ 0.3413$&$ 0.3189$&$$&$ 0.4132$&$ 0.4196$&$$&$ 0.3413$&$ 0.3195$\tabularnewline
&$-0.4186$&$-0.5234$&$$&$-0.3535$&$-0.4802$&$$&$-0.3242$&$-0.4542$&$$&$-0.3535$&$-0.4802$&$$&$-0.3242$&$-0.4539$\tabularnewline
occupation&$ 0.1033$&$ 0.0805$&$$&$ 0.0999$&$ 0.0742$&$$&$ 0.0999$&$ 0.0785$&$$&$ 0.1000$&$ 0.0741$&$$&$ 0.1000$&$ 0.0787$\tabularnewline
&$ 0.0759$&$ 0.0187$&$$&$ 0.0351$&$-0.0086$&$$&$ 0.0126$&$-0.0179$&$$&$ 0.0351$&$-0.0087$&$$&$ 0.0125$&$-0.0178$\tabularnewline
&$-0.0658$&$-0.1195$&$$&$-0.1104$&$-0.1501$&$$&$-0.1367$&$-0.1666$&$$&$-0.1104$&$-0.1502$&$$&$-0.1371$&$-0.1665$\tabularnewline
&$-1.2318$&$-1.4133$&$$&$-1.2430$&$-1.3711$&$$&$-1.2524$&$-1.3457$&$$&$-1.2433$&$-1.3709$&$$&$-1.2546$&$-1.3454$\tabularnewline
Distance&$ 0.8934$&\multicolumn{1}{c}{$-$}&$$&$ 0.6506$&$ 0.8409$&$$&$ 0.4547$&$ 0.8472$&$$&$ 0.6469$&$ 0.8357$&$$&$ 0.4484$&$ 0.8474$\tabularnewline
\multicolumn{15}{c}{} \\
Cutoff & \multicolumn{14}{c}{} \\
$\delta_1$&$-2.2134$&$-2.6098$&$$&$-2.0120$&$-1.8811$&$$&$-1.8218$&$-1.4920$&$$&$-2.0093$&$-1.8808$&$$&$-1.8210$&$-1.4854$\tabularnewline
$\delta_2$&$-0.2845$&$ 1.0077$&$$&$-0.0700$&$ 1.2409$&$$&$ 0.1428$&$ 1.4946$&$$&$-0.0667$&$ 1.2403$&$$&$ 0.1472$&$ 1.5034$\tabularnewline
$\delta_3$&$ 0.7932$&$ 2.2941$&$$&$ 0.9760$&$ 2.4528$&$$&$ 1.1711$&$ 2.6650$&$$&$ 0.9795$&$ 2.4522$&$$&$ 1.1762$&$ 2.6739$\tabularnewline
$\delta_4$&$ 2.3155$&$ 3.9963$&$$&$ 2.5117$&$ 4.1401$&$$&$ 2.7194$&$ 4.3499$&$$&$ 2.5154$&$ 4.1393$&$$&$ 2.7259$&$ 4.3586$\tabularnewline
Distance&$ 1.7262$&\multicolumn{1}{c}{$-$}&$$&$ 1.3650$&$ 1.6421$&$$&$ 1.0652$&$ 1.7066$&$$&$ 1.3590$&$ 1.6327$&$$&$ 1.0566$&$ 1.7151$\tabularnewline
\hline
\end{tabular}
\end{tabular}
}
\end{center}
\label{tb:result_affairs}
\end{table}%

\section{Conclusion and remarks}
\label{sec:conclude}
This study examined the problem of outliers in ordinal response model, which is a regression on ordered categorical data as the response variable.
\cite{scalera2021robust} derived the condition for the link functions to make the influence function in the maximum likelihood method bounded in the ordinal response model, but the probit link, logit link, complementary log-log link, and log-log link, which are commonly used, do not satisfy the condition.
Thus, when outliers exist in the data, the maximum likelihood method imposes restrictions on the link function, and that induces misspecification of the link function under such situations.
In our numerical experiments, we also confirmed that the misspecification of the link function induces a substantial bias in the estimation of the parameters.

\cite{scalera2021robust} mentions the boundedness of the influence function in the maximum likelihood method for the ordinal response model, and we further show that the influence function does not satisfy redescendence (Theorem \ref{thm:if} in Section \ref{sec:if}).
This means that the maximum likelihood method in the ordinal response model cannot completely eliminate the influence of large outliers, and in fact, in our numerical experiments, as the outlier ratio increases, the accuracy of the prediction becomes worse even if the cauchit link which has the bounded influence function is used. 

To solve these problems and to realize robust inference in the ordinal response model against outliers, we proposed inference methods in the ordinal response model using the DP and $\gamma$-divergences.
We also derived the influence functions for the methods with the DP and $\gamma$-divergences, and derived the conditions for the link functions to satisfy boundedness and redescendence (Theorems \ref{thm:dp_if} and \ref{thm:g_if} in Section \ref{sec:if}).
The commonly used link functions satisfy the conditions, and our method allows practitioners using ordinal response models to perform robust against outliers and flexible analysis.

Numerical experiments were conducted to evaluate the performance of our proposed methods with the DP and $\gamma$-divergences using bias, mean square error (MSE), and percentage of correctly classified responses (CCR).
In the absence of outliers, both methods perform as well as the maximum likelihood method, and in the presence of outliers, the results are almost unaffected by outliers.
As for the difference between the methods using the DP and $\gamma$-divergences, the bias value is slightly smaller when the DP divergence is used, but the MSE value is moderately larger than that using the $\gamma$-divergence.
In numerical experiments with two real data sets, the Boston housing and Affairs data, we also confirmed that our proposed methods with two robust divergences give the robust inference results that remove the influence of data that appear to be outliers.

\section*{Acknowledgement}
We thank Prof. Shounosuke Sugasawa and Prof. Kaoru Irie of the University of Tokyo for their very helpful comments.
This work was JSPS Grant-in-Aid for Early-Career Scientists Grant Number JP19K14597 and JSPS Grant-in-Aid for Scientific Research (B) Number 21H00699.

\def\thesection{Appendix}
\def\thesubsection{A.\arabic{subsection}}
\section{}
\setcounter{equation}{0}
\def\theequation{A.\arabic{equation}}
\def\thethm{A.\arabic{thm}}
\def\thelem{A.\arabic{lem}}


\subsection{Proof of Theorem \ref{thm:if}}
\label{app:ml_if}
First, consider the influence function of $\beta_k$ \eqref{eq:ml_b_if} for $k=1,2,\ldots,p$.
From appendix of \cite{scalera2021robust}, for the influence function of $\beta_k$ to have redescendence, the order of $u g'(u) / g(u)$ must be smaller than 1, where $g'(\cdot)$ is the first derivative of $g(\cdot)$.
This means that when the order of the tail of the function $g(u)$ is equal to $(\log u)^{-1}$, the influence function of $\beta_k$ has redescendence, however, no such distributions exist since distributions with order of tail more slowly than $u^{-1}$ are improper.

Next, consider the influence function of $\delta_l$ \eqref{eq:ml_d_if} for $l=1,2,\ldots,M-1$.
Since from Lagrange's theorem $G(\delta_{l}-\bm{x}_i^\top\bm{\beta}) - G(\delta_{l-1}-\bm{x}_i^\top\bm{\beta}) = (\delta_{l}-\delta_{l-1}) g(\delta^*-\bm{x}_i^\top\bm{\beta})$ for $\delta^*$ such that $\delta_{l-1} < \delta^* < \delta_{l}$ and $\delta_{l}-\delta_{l-1}>0$, for the influence function of $\delta_l$ to have redescendence, the order of $g(\delta_{l}-\bm{x}_i^\top\bm{\beta})/g(\delta^*-\bm{x}_i^\top\bm{\beta})$ must be $o(1)$.
However, such a distribution $G$ is improper, and such a distribution does not exist.

\qed

\subsection{Proof of Theorem \ref{thm:dp_if}}
\label{app:dp_if}
First, consider the influence function \eqref{eq:dp_b_if} of $\beta_k$.
It is clear that $\lim_{u\to\pm\infty}g(u)u = 0$ holds when the condition \eqref{eq:cod_dp_if} holds.
If here $u=\delta-\bm{x}_o^\top\bm{\beta}$ and note that $u$ goes to infinity of the same order as $x_{ok}$, since $G(\delta_{m}-\bm{x}_o^\top\bm{\beta}) - G(\delta_{m-1}-\bm{x}_o^\top\bm{\beta})$ is clearly bounded, the second term of equation \eqref{eq:dp_b_if} becomes zero for $u\to\pm\infty$.

Since from Lagrange's theorem, 
\begin{equation*}
G(\delta_{m}-\bm{x}_o^\top\bm{\beta}) - G(\delta_{m-1}-\bm{x}_o^\top\bm{\beta}) = (\delta_{m}-\delta_{m-1}) g(\delta^*-\bm{x}_o^\top\bm{\beta})
\end{equation*}
for $\delta^*$ such that $\delta_{m-1}<\delta^*<\delta_{m}$, then
\begin{align}
&g(\delta_{m}-\bm{x}_o^\top\bm{\beta}) \left[ G(\delta_{m}-\bm{x}_o^\top\bm{\beta}) - G(\delta_{m-1}-\bm{x}_o^\top\bm{\beta}) \right]^{\alpha-1} x_{ok} \nonumber \\
&= g(\delta_{m}-\bm{x}_o^\top\bm{\beta}) g(\delta^*-\bm{x}_o^\top\bm{\beta})^{\alpha-1} \frac{x_{ok}}{(\delta_{m}-\delta_{m-1})^{1-\alpha}} \label{eq:dp_if_pf}
\end{align}
and $g(\delta_{m}-\bm{x}_o^\top\bm{\beta}) g(\delta^*-\bm{x}_o^\top\bm{\beta})^{\alpha-1}$ is of the same order as $g(u)^{\alpha}$ for $u\to\pm\infty$.
Thus, if the condition \eqref{eq:cod_dp_if} holds, i.e., if there exists a certain $0<\alpha\leq1$ such that $\lim_{u\to\pm\infty} g(u)^\alpha u = 0$, then for $u\to\pm\infty$ the equation \eqref{eq:dp_if_pf} becomes
\begin{align*}
\lim_{u\to\pm\infty} g(u)^{\alpha} \frac{u}{(\delta_{m}-\delta_{m-1})^{1-\alpha}} = 0.
\end{align*}
Therefore, if the condition \eqref{eq:cod_dp_if} holds, the first term of equation \eqref{eq:dp_b_if} is zero for $u\to\pm\infty$ and then the influence function of $\beta_k$ using the DP divergence is also zero.

For the influence function \eqref{eq:dp_d_if} of $\delta_l$, it becomes zero for $u\to\pm\infty$ in the similar manner.
Namely, the influence functions \eqref{eq:dp_b_if} and \eqref{eq:dp_d_if} in the ordinal response model with the DP divergence are redescendent.

Moreover, the influence functions \eqref{eq:dp_b_if} and \eqref{eq:dp_d_if} are bounded since they are clearly continuous and their values are zero for $u\to\pm\infty$.

\qed

\bibliographystyle{myapalike2} 
\bibliography{References.bib}

\begin{thebibliography}{}

\bibitem[Agresti, 2010]{agresti2010analysis}
Agresti, A. (2010).
\newblock {\em Analysis of Ordinal Categorical Data}, volume 656.
\newblock John Wiley \& Sons.

\bibitem[Agresti and Kateri, 2017]{agresti2017ordinal}
Agresti, A. and Kateri, M. (2017).
\newblock Ordinal probability effect measures for group comparisons in
  multinomial cumulative link models.
\newblock {\em Biometrics}, {\bfseries 73}(1), 214--219.

\bibitem[Albert and Chib, 1993]{albert1993bayesian}
Albert, J.~H. and Chib, S. (1993).
\newblock Bayesian analysis of binary and polychotomous response data.
\newblock {\em Journal of the American Statistical Association}, {\bfseries
  88}(422), 669--679.

\bibitem[Ashby\textit{ et~al.}, 1989]{ashby1989ordered}
Ashby, D., West, C.~R., and Ames, D. (1989).
\newblock The ordered logistic regression model in psychiatry: rising
  prevalence of dementia in old people's homes.
\newblock {\em Statistics in Medicine}, {\bfseries 8}(11), 1317--1326.

\bibitem[Baetschmann\textit{ et~al.}, 2020]{baetschmann2020feologit}
Baetschmann, G., Ballantyne, A., Staub, K.~E., and Winkelmann, R. (2020).
\newblock feologit: A new command for fitting fixed-effects ordered logit
  models.
\newblock {\em The Stata Journal}, {\bfseries 20}(2), 253--275.

\bibitem[Basak\textit{ et~al.}, 2021]{basak2021optimal}
Basak, S., Basu, A., and Jones, M. (2021).
\newblock On the 'optimal' density power divergence tuning parameter.
\newblock {\em Journal of Applied Statistics}, {\bfseries 48}(3), 536--556.

\bibitem[Basu\textit{ et~al.}, 1998]{basu1998robust}
Basu, A., Harris, I.~R., Hjort, N.~L., and Jones, M. (1998).
\newblock Robust and efficient estimation by minimising a density power
  divergence.
\newblock {\em Biometrika}, {\bfseries 85}(3), 549--559.

\bibitem[Breiger, 1981]{breiger1981social}
Breiger, R.~L. (1981).
\newblock The social class structure of occupational mobility.
\newblock {\em American Journal of Sociology}, {\bfseries 87}(3), 578--611.

\bibitem[Castilla\textit{ et~al.}, 2021]{castilla2021estimation}
Castilla, E., Jaenada, M., and Pardo, L. (2021).
\newblock {Estimation and testing on independent not identically distributed
  observations based on R\'enyi's pseudodistances}.
\newblock {\em arXiv preprint arXiv:2102.12282}, {\bfseries }.

\bibitem[Christensen, 2018]{christensen2018cumulative}
Christensen, R. H.~B. (2018).
\newblock Cumulative link models for ordinal regression with the r package
  ordinal.
\newblock {\em Submitted in Journal of Statistical Software}, {\bfseries 35}.

\bibitem[Croux\textit{ et~al.}, 2013]{croux2013robust}
Croux, C., Haesbroeck, G., and Ruwet, C. (2013).
\newblock Robust estimation for ordinal regression.
\newblock {\em Journal of Statistical Planning and Inference}, {\bfseries
  143}(9), 1486--1499.

\bibitem[Czado and Santner, 1992]{czado1992effect}
Czado, C. and Santner, T.~J. (1992).
\newblock The effect of link misspecification on binary regression inference.
\newblock {\em Journal of Statistical Planning and Inference}, {\bfseries
  33}(2), 213--231.

\bibitem[Dawid and Musio, 2015]{dawid2015bayesian}
Dawid, A.~P. and Musio, M. (2015).
\newblock Bayesian model selection based on proper scoring rules.
\newblock {\em Bayesian Analysis}, {\bfseries 10}(2), 479--499.

\bibitem[Desgagn{\'e}, 2015]{desgagne2015robustness}
Desgagn{\'e}, A. (2015).
\newblock Robustness to outliers in location--scale parameter model using
  log-regularly varying distributions.
\newblock {\em The Annals of Statistics}, {\bfseries 43}(4), 1568--1595.

\bibitem[Fair, 1978]{fair1978theory}
Fair, R.~C. (1978).
\newblock A theory of extramarital affairs.
\newblock {\em Journal of Political Economy}, {\bfseries 86}(1), 45--61.

\bibitem[Franses and Paap, 2001]{franses2001quantitative}
Franses, P.~H. and Paap, R. (2001).
\newblock {\em Quantitative Models in Marketing Research}.
\newblock Cambridge University Press.

\bibitem[Fujisawa and Eguchi, 2008]{fujisawa2008robust}
Fujisawa, H. and Eguchi, S. (2008).
\newblock Robust parameter estimation with a small bias against heavy
  contamination.
\newblock {\em Journal of Multivariate Analysis}, {\bfseries 99}(9),
  2053--2081.

\bibitem[Gagnon\textit{ et~al.}, 2020]{gagnon2020new}
Gagnon, P., Desgagn{\'e}, A., and B{\'e}dard, M. (2020).
\newblock A new bayesian approach to robustness against outliers in linear
  regression.
\newblock {\em Bayesian Analysis}, {\bfseries 15}(2), 389--414.

\bibitem[Ghosh and Basu, 2013]{ghosh2013robust}
Ghosh, A. and Basu, A. (2013).
\newblock Robust estimation for independent non-homogeneous observations using
  density power divergence with applications to linear regression.
\newblock {\em Electronic Journal of Statistics}, {\bfseries 7}, 2420--2456.

\bibitem[Hampel, 1974]{hampel1974influence}
Hampel, F.~R. (1974).
\newblock The influence curve and its role in robust estimation.
\newblock {\em Journal of the American Statistical Association}, {\bfseries
  69}(346), 383--393.

\bibitem[Hampel\textit{ et~al.}, 1986]{hampel1986robust}
Hampel, F.~R., Ronchetti, E.~M., Rousseeuw, P.~J., and Stahel, W.~A. (1986).
\newblock {\em Robust Statistics}.
\newblock Wiley Online Library.

\bibitem[Hamura\textit{ et~al.}, 2022]{hamura2022log}
Hamura, Y., Irie, K., and Sugasawa, S. (2022).
\newblock Log-regularly varying scale mixture of normals for robust regression.
\newblock {\em Computational Statistics and Data Analysis}, {\bfseries 173},
  107517.

\bibitem[Harrison and Rubinfeld, 1978]{harrison1978hedonic}
Harrison, J.~D. and Rubinfeld, D.~L. (1978).
\newblock Hedonic housing prices and the demand for clean air.
\newblock {\em Journal of Environmental Economics and Management}, {\bfseries
  5}(1), 81--102.

\bibitem[Hashimoto and Sugasawa, 2020]{hashimoto2020robust}
Hashimoto, S. and Sugasawa, S. (2020).
\newblock Robust bayesian regression with synthetic posterior distributions.
\newblock {\em Entropy}, {\bfseries 22}(6), 661.

\bibitem[Huber and Ronchetti, 2009]{huber2009robust}
Huber, J. and Ronchetti, E.~M. (2009).
\newblock {\em Robust Statistics, Second Edition}.
\newblock Wiley.

\bibitem[Iannario\textit{ et~al.}, 2017]{iannario2017robust}
Iannario, M., Monti, A.~C., Piccolo, D., and Ronchetti, E. (2017).
\newblock Robust inference for ordinal response models.
\newblock {\em Electronic Journal of Statistics}, {\bfseries 11}(2),
  3407--3445.

\bibitem[Jones\textit{ et~al.}, 2001]{jones2001comparison}
Jones, M., Hjort, N.~L., Harris, I.~R., and Basu, A. (2001).
\newblock A comparison of related density-based minimum divergence estimators.
\newblock {\em Biometrika}, {\bfseries 88}(3), 865--873.

\bibitem[Kawakami and Hashimoto, 2022]{kawakami2022approximate}
Kawakami, J. and Hashimoto, S. (2022).
\newblock Approximate gibbs sampler for bayesian huberized lasso.
\newblock {\em arXiv preprint arXiv:2204.00237}, {\bfseries }.

\bibitem[Kawashima and Fujisawa, 2017]{kawashima2017robust}
Kawashima, T. and Fujisawa, H. (2017).
\newblock Robust and sparse regression via $\gamma$-divergence.
\newblock {\em Entropy}, {\bfseries 19}(11), 608.

\bibitem[Maronna\textit{ et~al.}, 2019]{maronna2019robust}
Maronna, R.~A., Martin, R.~D., Yohai, V.~J., and Salibi{\'a}n-Barrera, M.
  (2019).
\newblock {\em Robust Statistics: Theory and Methods (with R)}.
\newblock John Wiley \& Sons.

\bibitem[McCullagh, 1980]{mccullagh1980regression}
McCullagh, P. (1980).
\newblock Regression models for ordinal data.
\newblock {\em Journal of the Royal Statistical Society: Series B
  (Methodological)}, {\bfseries 42}(2), 109--127.

\bibitem[O'Hagan and Pericchi, 2012]{o2012bayesian}
O'Hagan, A. and Pericchi, L. (2012).
\newblock Bayesian heavy-tailed models and conflict resolution: A review.
\newblock {\em Brazilian Journal of Probability and Statistics}, {\bfseries
  26}(4), 372--401.

\bibitem[Pyne\textit{ et~al.}, 2022]{pyne2022robust}
Pyne, A., Roy, S., Ghosh, A., and Basu, A. (2022).
\newblock Robust and efficient estimation in ordinal response models using the
  density power divergence.
\newblock {\em arXiv preprint arXiv:2208.14011}, {\bfseries }.

\bibitem[Riani\textit{ et~al.}, 2011]{riani2011outliers}
Riani, M., Torti, F., and Zani, S. (2011).
\newblock Outliers and robustness for ordinal data.
\newblock {\em Modern Analysis of Customer Surveys: with applications using R},
  {\bfseries }, 155--169.

\bibitem[Scalera\textit{ et~al.}, 2021]{scalera2021robust}
Scalera, V., Iannario, M., and Monti, A.~C. (2021).
\newblock Robust link functions.
\newblock {\em Statistics}, {\bfseries 55}(4), 963--977.

\bibitem[Shao\textit{ et~al.}, 2019]{shao2019bayesian}
Shao, S., Jacob, P.~E., Ding, J., and Tarokh, V. (2019).
\newblock Bayesian model comparison with the hyv{\"a}rinen score: Computation
  and consistency.
\newblock {\em Journal of the American Statistical Association}, {\bfseries
  114}, 1826--1837.

\bibitem[Sugasawa and Yonekura, 2021]{sugasawa2021selection}
Sugasawa, S. and Yonekura, S. (2021).
\newblock On selection criteria for the tuning parameter in robust divergence.
\newblock {\em Entropy}, {\bfseries 23}(9), 1147.

\bibitem[Tomizawa\textit{ et~al.}, 2006]{tomizawa2006decompositions}
Tomizawa, S., Miyamoto, N., and Ouchi, M. (2006).
\newblock Decompositions of symmetry model into marginal homogeneity and
  distance subsymmetry in square contingency tables with ordered categories.
\newblock {\em REVSTAT--Statistical Journal}, {\bfseries 4}(2), 153--161.

\bibitem[Uebersax, 1999]{uebersax1999probit}
Uebersax, J.~S. (1999).
\newblock Probit latent class analysis with dichotomous or ordered category
  measures: Conditional independence/dependence models.
\newblock {\em Applied Psychological Measurement}, {\bfseries 23}(4), 283--297.

\bibitem[Warwick and Jones, 2005]{warwick2005choosing}
Warwick, J. and Jones, M. (2005).
\newblock Choosing a robustness tuning parameter.
\newblock {\em Journal of Statistical Computation and Simulation}, {\bfseries
  75}(7), 581--588.

\end{thebibliography}
\end{document}